\documentclass[12pt]{iopart}

\usepackage{iopams}
\usepackage{graphicx}
\usepackage[T1]{fontenc}
\usepackage{color}
\usepackage{textcomp}
\usepackage{hyperref}

\usepackage{lineno}

\usepackage{geometry}
\begin{document}	
\noindent {\small Accepted manuscript for publication in Materials for Quantum Technology (CC BY 4.0).\\
\url{https://doi.org/10.1088/2633-4356/abcbeb}\\}
\rule{\linewidth}{0.05cm}

\title[Phonon Assisted Excitation]{Controlling Photoluminescence Spectra of hBN Color Centers by Selective Phonon-Assisted Excitation: A Theoretical Proposal}

\author{Daniel~Groll$^{1}$, Thilo~Hahn$^1$, Pawe\l{}~Machnikowski$^2$, Daniel~Wigger$^{2}$ and Tilmann~Kuhn$^1$}

\address{$^1$Institute of Solid State Theory, University of M\"{u}nster, 48149 M\"{u}nster, Germany}
\address{$^2$Department of Theoretical Physics, Wroc\l{}aw University of Science and Technology, 50-370 Wroc\l{}aw, Poland }

\ead{daniel.groll@uni-muenster.de, tilmann.kuhn@uni-muenster.de}

\begin{abstract}
Color centers in hexagonal boron nitride (hBN) show stable single photon emission even at room temperature, making these systems a promising candidate for quantum information applications. Besides this remarkable property, also their interaction with longitudinal optical (LO) phonons is quite unique because they lead to dominant phonon sidebands (PSBs), well separated from the zero phonon line (ZPL). In this work we utilize this clear spectral separation to theoretically investigate the influence of phonon decay dynamics on time-dependent photoluminescence (PL) signals. Our simulations show, that by using tailored optical excitation schemes it is possible to create a superposition between the two LO modes, leading to a phonon quantum beat that manifests in the time-dependent PL signal.
\end{abstract}


\section{Introduction}
A few years ago the family of solid state single photon emitters (SPEs) was joined by a promising new group in the form of color centers in hexagonal boron nitride (hBN)~\cite{tran2016quaI}. These systems show single photon emission even at room temperature~\cite{tran2016quaI, tran2016quaII, tran2016rob, martinez2016eff, exarhos2017opt}. Despite rapid intense investigations the exact atomic structure is still under debate~\cite{tran2016quaI, tran2016rob, tawfik2017fir,sajid2018def, abdi2018col, gottscholl2020initialization, sajid2020single}. However a few candidates have been identified to likely contain carbon~\cite{tawfik2017fir,sajid2018def,mendelson2020identifying}. Beyond its characterization, recent studies have focused on the coherent optical control~\cite{konthasinghe2019rabi} and the manipulation of spin states with external magnetic fields~\cite{gottscholl2020initialization,kianinia2020generation}.\\
All solid state SPEs unavoidably interact with lattice distortions and therefore couple to phonons~\cite{grosso2017tun,mendelson2020strain}. In this context many former studies have focused on minimizing the impact of the phonon coupling on SPEs~\cite{axt2005reducing, grange2017reducing}. This could in principle be possible in hBN color centers by reducing the dipole moment of the emitter wave function~\cite{wigger2019phonon}. Here, we follow a different approach by making direct use of this interaction to control the emitter by tailored phonon-assisted excitations.\\
The coupling between phonons and the hBN-SPEs has some unique properties, compared to other emitters, like semiconductor quantum dots (QDs) or NV centers in diamond. In QDs the exciton-phonon interaction is rather weak, leading to emission spectra that are dominated by the zero phonon line (ZPL)~\cite{besombes2001aco,borri2001ultralong,krummheuer2002the,forstner2003phonon,jakubczyk2016imp}. The QD excitons couple predominantly to acoustic phonons~\cite{krummheuer2002the}, which often render a limiting factor to coherent control~\cite{forstner2003phonon,axt2005reducing,ramsay2010damping,jakubczyk2016imp}. In NV centers in diamond the optical response is dominated by phonon sidebands (PSBs) which mainly stem from local mode oscillations with energies of a few ten meVs~\cite{zaitsev2000vib,huxter2013vib,kehayias2013inf}. Especially at room temperature the PSBs strongly overlap and make a separation from the ZPL challenging~\cite{jelezko2001spectroscopy,beha2012opt}. Compared to these two types of SPEs the phonon coupling of hBN color centers is quite unique. Because of the strong polarity of the hBN crystal and the color centers therein coupling to longitudinal optical (LO) modes is remarkably strong. It results in peaks that are well separated from the ZPL by approximately 200 meV and that can reach heights of up to half the ZPL~\cite{wigger2019phonon}.\\
It has been shown that optical excitation of SPEs via phonon-assisted processes is very efficient. While in QDs typically excitation via acoustic PSBs is chosen~\cite{glassl2013proposed, reiter2014role, quilter2015phonon}, in hBN the LO-PSBs provide an efficient excitation channel~\cite{wigger2019phonon, khatri2020optical}. This makes these modes excellent candidates to investigate the underlying interplay with the color center in more detail and study the influence of phonon dynamics. So far only equilibrium spectra of color centers in hBN have been studied in detail in the literature~\cite{exarhos2017opt,wigger2019phonon,grosso2020low}. Here, we go a step further and simulate time-dependent photoluminescence (PL) spectroscopy signals. We model an actual experimental realization of this dynamically and spectrally resolved PL measurement. There we focus on the phonons' effect on non-equilibrium dynamics of PL spectra after a pulsed laser excitation. In practice we study the thermalization process and quantum beats of the LO phonons. This is a first important step towards more advanced quantum state preparation of the phonons by tailored optical excitation schemes of the color center~\cite{sauer2010lattice,chu2018creation,hahn2019influence}.
\section{Theory}
In this work we calculate time-dependent PL spectra of a hBN color center after a phonon-assisted excitation with a short laser pulse. The color center inevitably interacts with the LO phonons of the host lattice. Further dephasing and decay of the emitter, as well as decay of the LO modes due to anharmonicities, are accounted for by Lindblad dissipators.
\subsection{The model}
\begin{figure}[t]
	\centering
	\includegraphics[width=0.25\columnwidth]{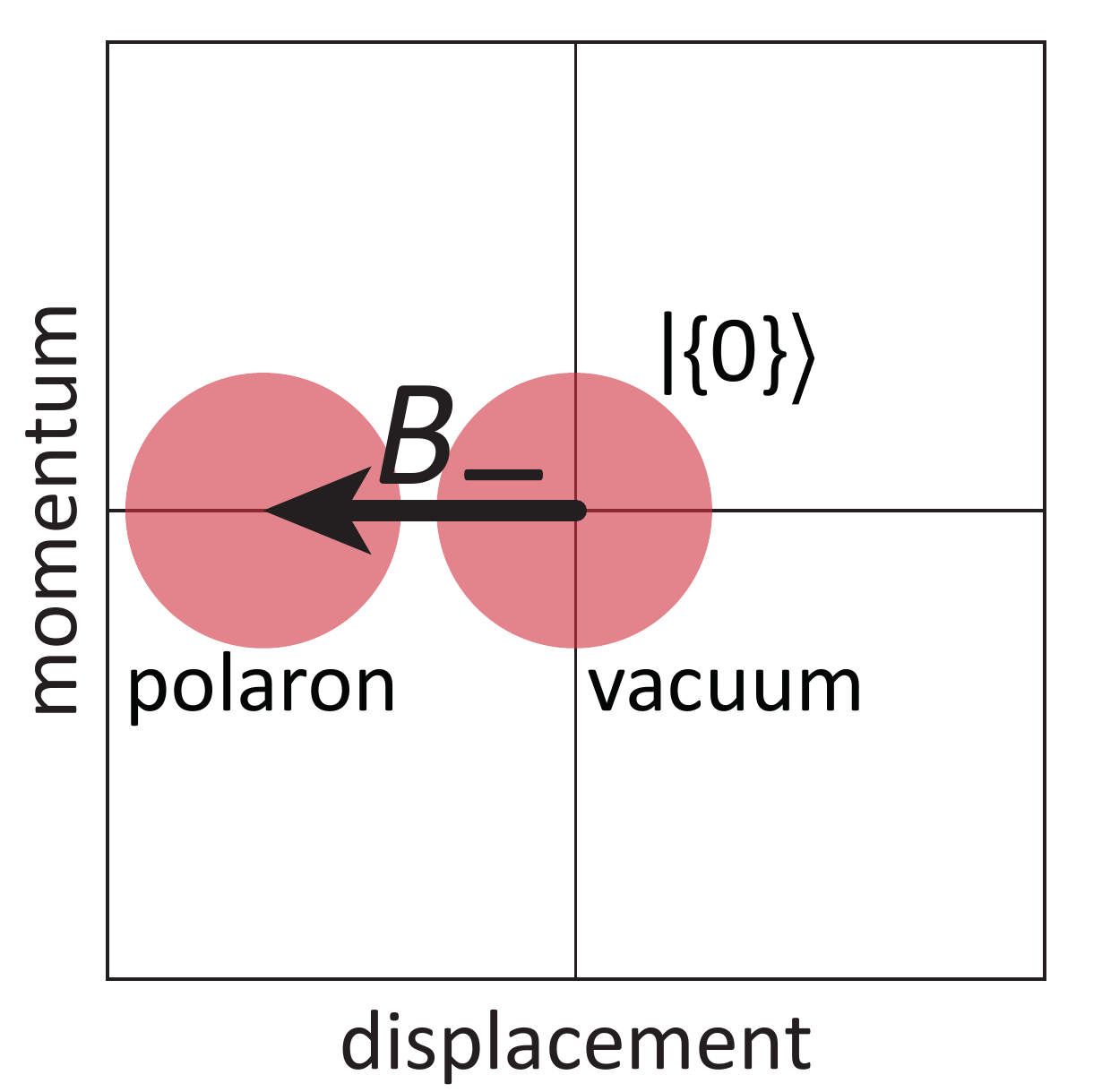}
	\caption{Schematic picture of the phononic phase space of a single mode. The excitation of the TLS results in a new equilibrium for the phonons called polaron. The polaron state is shifted in phase space with respect to the vacuum by $B_-$.}
	\label{fig:0}
\end{figure}

For the states of the hBN color center we assume a two-level system (TLS) consisting of a ground state $\left| G\right>$ and an excited state $\left| X\right>$. Although some studies suggest additional dark states that lie below the optically active one~\cite{tran2016quaI,martinez2016eff,kianinia2018all}, we restrict our model to the TLS for the following reasons. It was shown that this approach allows to reproduce optical spectra~\cite{wigger2019phonon} and we are here only interested in (sub-)picosecond dynamics, which is much faster than typical transitions involving the dark states~\cite{kianinia2018all}.\\
In Ref.~\cite{wigger2019phonon} it was shown that the independent boson model is suitable to calculate absorption spectra of single color centers. Therefore we start with the Hamiltonian~\cite{mahan2013many,krummheuer2002the}
\begin{eqnarray}\label{eq:H_0}
H_0&=&\hbar \omega_X X^{\dagger}X+\sum_j \hbar\omega_j^{} b_j^{\dagger}b_j^{} \nonumber\\
&&+X^{\dagger}X\sum_j \left( \hbar g_j^{}b_j^{\dagger}+\hbar g_j^*b_j^{} \right)\,,
\end{eqnarray}
where $X=\left|G\right>\left<X\right|$ is the transition operator from the excited to the ground state with an energy splitting $\hbar\omega_X$ that is typically in the optical range. $\omega_j^{}$ is the frequency of the $j$-th phonon mode with annihilation and creation operators $b_j^{}$ and $b_j^{\dagger}$, respectively. The coupling constants are given by $g_j^{}$, where the index $j$ collectively denotes all quantum numbers, that are used to label the phonon modes, i.e., branch index and momentum. Low energy phonons like acoustic and local modes are hard to separate from the ZPL contribution at room temperature because of their significant spectral overlap. But the LO modes in hBN have energies of up to 200~meV, which makes them easy to isolate spectrally even at room temperature. For simplicity we assume dispersionless LO branches, which already led to excellent results in Ref.~\cite{wigger2019phonon}. In this case, we can describe a single LO branch by a single collective boson mode \cite{stauber2000electron}.\\
The coupling of the emitter to the electrical field of a laser pulse is described within the usual dipole and rotating wave approximation via the interaction Hamiltonian
\numparts
\begin{equation}
H_I(t)=\frac12 \left[\mathcal{E}(t)^*X+\mathcal{E}(t)X^{\dagger}\right] \,,
\end{equation}
where the effective laser field reads
\begin{eqnarray}
\mathcal{E}(t)&=&-2 \left<X\right| \boldsymbol{d}\cdot\boldsymbol{E}_0(t) \left|G\right> \exp(-i\omega_l t)\nonumber\\
&=&\frac{\hbar\mathcal{E}_0}{\sqrt{2\pi\sigma^2}}\exp\left(-\frac{t^2}{2\sigma^2}-i\omega_l t\right)\,
\end{eqnarray}
\endnumparts
and $\boldsymbol{d}$ and $\boldsymbol{E}_0$ are the dipole operator and the real envelope of the electrical field of the laser pulse, respectively. We choose the effective laser field to be a Gaussian pulse with temporal width $\sigma$, centered around time $t=0$, with the dimensionless pulse area $\mathcal{E}_0$.\\
The full unitary dynamics of the system is described by the Hamiltonian $H(t)=H_0+H_I(t)$. In absence of the laser field, the Hamiltonian reduces to $H_0$, which can be diagonalized by the so-called polaron transformation~\cite{mahan2013many, alicki2004pure, roszak2006path, nazir2016modelling}.  With this transformation one can find the eigenstates of $H_0$ to be
\numparts
\begin{equation}
\left|G\right> \otimes \left|\{n\}\right>\,,\quad \left|X\right> \otimes B_- \left|\{n\}\right>\,,
\end{equation}
where $\left|\{n\}\right>$ is an arbitrary multi-mode Fock state and
\begin{equation}\label{eq:B_pm}
B_{\pm}=\exp\left[\pm\sum_j \frac{1}{\omega_j}(g_j^{}b_j^{\dagger}-g_j^*b_j^{})\right]\,
\end{equation}
\endnumparts
are multi-mode displacement operators, displacing each phonon mode by $\pm g_j^{}/\omega_j^{}$ in phase space. We see that the ground state of the phonon subsystem is the multi-mode vacuum $\left|\{0\}\right>$, in the case that the TLS is in the ground state $\left|G\right>$. However for the case, that it is in the excited state $\left|X\right>$, the ground state for the phonon subsystem is the displaced vacuum $B_- \left|\{0\}\right>$. As schematically depicted in Fig.~\ref{fig:0} this simply describes that, induced by an excitation of the TLS, the lattice relaxes to a new equilibrium position forming the so-called polaron~\cite{wigger2014energy, wigger2020acoustic}. It also leads to a renormalization of the transition energy by the polaron-shift to $\hbar \tilde{\omega}_X=\hbar \omega_X-\sum_j\hbar |g_j^{}|^2/\omega_j^{}$. Consequently also the coupling to the laser field is renormalized, such that the pulse area changes as $\mathcal{E}_0\rightarrow B\mathcal{E}_0$, where $B$ is the expectation value of $B_{\pm}$ with respect to the initial thermal phonon state~\cite{krugel2005role, nazir2016modelling}. Note, that $B$ is the same for $B_+$ and $B_-$ because it is just given by the overlap between the displaced and the original thermal distribution in phase space. Therefore it is independent of the direction of the displacement.
\subsection{Dissipation}
We describe the non-unitary dynamics of the system by a set of Lindblad dissipators, i.e., we restrict ourselves to Markovian dynamics~\cite{breuer2002theory}. Decay and pure dephasing of the TLS are described by the dissipators~\cite{roy2011influence}
\numparts
\begin{equation}
\mathcal{D}_{\mathrm{xd}}(\rho)=\frac{1}{T_{\mathrm{xd}}}\left( X\rho X^{\dagger}-\frac{1}{2}\left\{X^{\dagger}X,\rho\right\} \right)
\end{equation}
and 
\begin{equation}
\mathcal{D}_{\mathrm{pd}}(\rho)=\frac{2}{T_{\mathrm{pd}}}\left( X^{\dagger}X\rho X^{\dagger}X-\frac{1}{2}\left\{X^{\dagger}X,\rho\right\} \right)\,,
\end{equation}
\endnumparts
respectively, acting on the density matrix $\rho$. Here, $T_{\mathrm{xd}}$ is the excited state decay time, i.e., the lifetime of the emitter and $T_{\mathrm{pd}}$ the additional pure dephasing time. Note that the decay automatically includes a dephasing time $2T_{\mathrm{xd}}$. It is known that the lifetime of the hBN emitters is in the nanosecond range independent of temperature while the detected dephasing time is much shorter. Therefore the additional pure dephasing has to be considered here.\\
We will later focus on the dynamics of the high energy LO modes. For these modes two aspects are important: (A) Due to their energies way above 100~meV they are thermally unoccupied even at room temperature and (B) they typically have lifetimes of a few ps \cite{cusco2018isotopic}. To account for (B) we use a dissipator, describing the decay of a single LO mode $j$ with the rate $\gamma_j$, given by
\begin{eqnarray}
\mathcal{D}_{{\rm LO}_j}(\rho)&=& \mathcal{D}_{{\rm LO}_j}^{G}(\rho)+\mathcal{D}_{{\rm LO}_j}^{X}(\rho)+\mathcal{D}_{{\rm LO}_j}^{XG}(\rho) \label{eq:D_LO}\,, \\
\mathcal{D}_{{\rm LO}_j}^{G}(\rho)&=&\gamma_j^{}\Big( \left|G\right>\left<G\right| b_j^{} \rho b_j^{\dagger} \left|G\right>\left<G\right| \nonumber\\
&&\qquad -\frac{1}{2}\left\{ \left|G\right> \left<G\right| b_j^{\dagger}b_j^{}, \rho\right\} \Big) \nonumber\,,\\
\mathcal{D}_{{\rm LO}_j}^{X}(\rho)&=&\gamma_j^{} \Big( \left|X\right> \left<X\right| B_- b_j^{} B_+ \rho B_- b_j^{\dagger} B_+ \left|X\right>\left<X\right| \nonumber\\
&&\qquad -\frac{1}{2} \left\{\left|X\right> \left<X\right| B_- b_j^{\dagger} b_j^{} B_+, \rho\right\}\Big)\,,\nonumber\\
\mathcal{D}_{{\rm LO}_j}^{XG}(\rho)&=&\gamma_j \Big(\left|G\right>\left<G\right|b_j\rho B_-b_j^{\dagger}B_+\left|X\right>\left<X\right|\nonumber\\
&&\qquad+\left|X\right>\left<X\right|B_-b_jB_+\rho b_j^{\dagger}\left|G\right>\left<G\right|\Big)\nonumber\,.
\end{eqnarray}
The dissipator from Eq.~\eref{eq:D_LO} consists of three parts: $\mathcal{D}_{{\rm LO}_j}^{G}(\rho)$ describes the relaxation of the mode LO$_j$ to its vacuum, in the case that the emitter is in the ground state $\left| G\right>$. $\mathcal{D}_{{\rm LO}_j}^{X}(\rho)$ is more involved and describes the relaxation into the displaced vacuum $B_-\left|\{0\}\right>$, i.e., the polaron configuration, when the emitter is in the excited state $\left|X\right>$. $\mathcal{D}_{{\rm LO}_j}^{XG}(\rho)$ does not lend itself to such an easy interpretation, but we see that it only acts on the polarization of the TLS, i.e., density matrix elements proportional to $\left|G\right>\left<X\right|$ or $\left|X\right>\left<G\right|$, leaving the state of the TLS unchanged. As discussed in the next section, the PL spectrum is determined only by the occupation of the TLS, i.e., density matrix elements proportional to $\left|X\right>\left<X\right|$, apart from the time during the laser pulse interaction. Since the laser pulse duration that is chosen in this work is $\sigma=80$~fs, much shorter than typical values of the LO phonon lifetime~\cite{cusco2018isotopic}, this part of the dissipator is of minor importance. This has been checked by comparing numerical calculations of spectra with and without $\mathcal{D}_{{\rm LO}_j}^{XG}(\rho)$. The derivation of Eq.~\eref{eq:D_LO} is given in the Supplementary Material (SM). Finally, the Lindblad equation that governs the dynamics of the system is given by
\begin{eqnarray}\label{eq:Lindblad}
\frac{\rm d}{{\rm d}t}\rho(t)&=&-\frac{i}{\hbar}\left[ H(t),\rho(t)\right]\\
&&+\mathcal{D}_{\mathrm{xd}}[\rho(t)]+\mathcal{D}_{\mathrm{pd}}[\rho(t)]+\sum_j \mathcal{D}_{{\rm LO}_j}[\rho(t)]\,. \nonumber
\end{eqnarray}
\subsection{Time-dependent spectroscopy}
To model time-dependent PL spectra, we consider a filter, e.g. a Fabry-Perot interferometer, that selects the light emitted from the hBN defect, before the intensity of the filtered light field is detected as a function of time. The spectral width of the filter will at the same time yield a finite temporal resolution due to time-energy uncertainty. The time-dependent PL spectrum of the emitter, using a filter with a Lorentzian spectrum, is given by~\cite{mollow1969power,eberly1977time}
\begin{eqnarray}\label{eq:spectrum}
S(t,\omega;\Gamma)&=&2\Gamma^2 {\rm Re}\Bigg[\int\limits_{0}^{\infty}{\rm d}\tau \int\limits_{-\infty}^t {\rm d} t'e^{-\Gamma\tau}e^{-2\Gamma (t-t')}e^{-i\omega\tau} \nonumber\\
&&\quad\quad\quad\qquad\qquad\times \left< X^{\dagger}(t')X(t'-\tau)\right> \Bigg] \,.
\end{eqnarray}
A short derivation of this formula is given in the SM. We see, that the spectral width $\Gamma$ of the spectrometer leads to a finite spectral resolution, due to its appearance in the $\tau$-integration. At the same time it appears as a finite integration time $\Gamma^{-1}$ over the history of the system up until time $t$, which results in the finite time resolution mentioned before. Consequently, a perfect spectral and temporal resolution cannot be achieved at the same time. By integrating Eq.~\eref{eq:spectrum} over the time $t$, we obtain the (time-integrated) PL spectrum of the emitter. The central quantity to determine the time-dependent PL spectrum is the causal two-time correlation function $\left< X^{\dagger}(t)X(t-\tau)\right>$ of the TLS. We will calculate this quantity numerically, applying the quantum regression theorem, together with Eq.~\eref{eq:Lindblad}~\cite{breuer2002theory}. The regression theorem can be applied here because the dynamics of our TLS+LO phonon system is described by a Lindblad equation and is therefore Markovian (see SM for more details). We perform the calculations in a truncated Hilbert space limiting the maximum number of phonons per mode. An important property of this correlation function is that, if $t$ and $t-\tau$ are times after the interaction with the laser pulse, only the part of the density matrix proportional to $\left|X\right>\left<X\right|$ contributes to the PL signal, since $H_0$ does not lead to any transitions in the TLS. This is true even in the presence of the dissipators in Eq.~\eref{eq:Lindblad}, as discussed in the SM. Thus, the subspace of the system, in which the state $\left|X\right>$ is occupied, entirely determines the spectrum, apart from contributions during the laser pulse interaction.
\section{Results}
Before coming to the characteristic two--LO-mode structure of hBN color centers we simplify the system and consider just a single LO mode. This {\it reduced model} allows us to instructively explain the influence of a phonon-assisted excitation and the decay of the LO phonon mode on the PL signal dynamics. We furthermore neglect the influence of low energy phonons like acoustic modes. Therefore we will not get an asymmetric broadening of the ZPL and additional broadenings of the PSBs. However, since all phonon modes couple independently to the TLS, this will not influence the dynamics induced by the LO modes significantly. Due to the possible LO energies between $165$~meV and $200$~meV~\cite{wigger2019phonon} we can safely assume the thermal equilibrium state for these modes to be the vacuum state $\left|G\right>\otimes\left|\{0\}\right>$ even at room temperature.\\
In the following we choose the ZPL energy to be $\hbar\tilde{\omega}_X=2$~eV. While the lifetime of the emitter $T_{\mathrm{xd}}=4$~ns is considered to be independent of the temperature, the dephasing time changes with temperature and we choose $T_{\mathrm{pd}}=80$~fs at 300~K, $T_{\mathrm{pd}}=400$~fs at 70~K, and $T_{\mathrm{pd}}=800$~fs at 4~K~\cite{wigger2019phonon}. This reported temperature dependence might originate from nonlinear phonon coupling effects~\cite{machnikowski2006change} or spectral diffusion induced by charge fluctuations~\cite{spokoyny2020effect}. The LO phonon lifetime is typically of a few ps~\cite{cusco2018isotopic} and we choose $\gamma_j^{-1}=2$~ps for all LO modes at all temperatures. In principle the phonon lifetime arising from anharmonicities should also depend on temperature. However, for the sake of clarity and due to lack of experimental evidence we consider a fixed LO phonon lifetime. It is important to note that the qualitative behavior of the system does not depend on the exact numerical value of the phonon lifetime, as long as its order of magnitude is correct. For the spectrometer we choose a resolution of $\hbar\Gamma=10$~meV, which corresponds to a temporal resolution of $\Gamma^{-1}\approx 66$~fs. We will excite the system with a pulse of duration $\sigma=80$~fs, i.e., a spectral width of $\hbar/\sigma\approx 8.2$~meV. The renormalized pulse area is set to $B \mathcal{E}_0=0.2\approx0.06\pi$, well within the linear response regime.
\subsection{Single LO mode - reduced model}
We start the discussion with the simplified model of a single phonon mode coupling to the emitter. In this case the sums over all phonon modes $\sum_j \dots$ in Eqs.~\eref{eq:H_0}, \eref{eq:B_pm}, and \eref{eq:Lindblad} reduce to a single term. For the energy of this single mode we choose the average of the two modes in hBN, i.e., $\hbar\omega_{\rm LO}=182.5$~meV. The effective coupling strength to the emitter is chosen to be $g_{\rm LO}=0.5\,\omega_{\rm LO}$.
\begin{figure}[h]
	\centering
	\includegraphics[width=0.5\columnwidth]{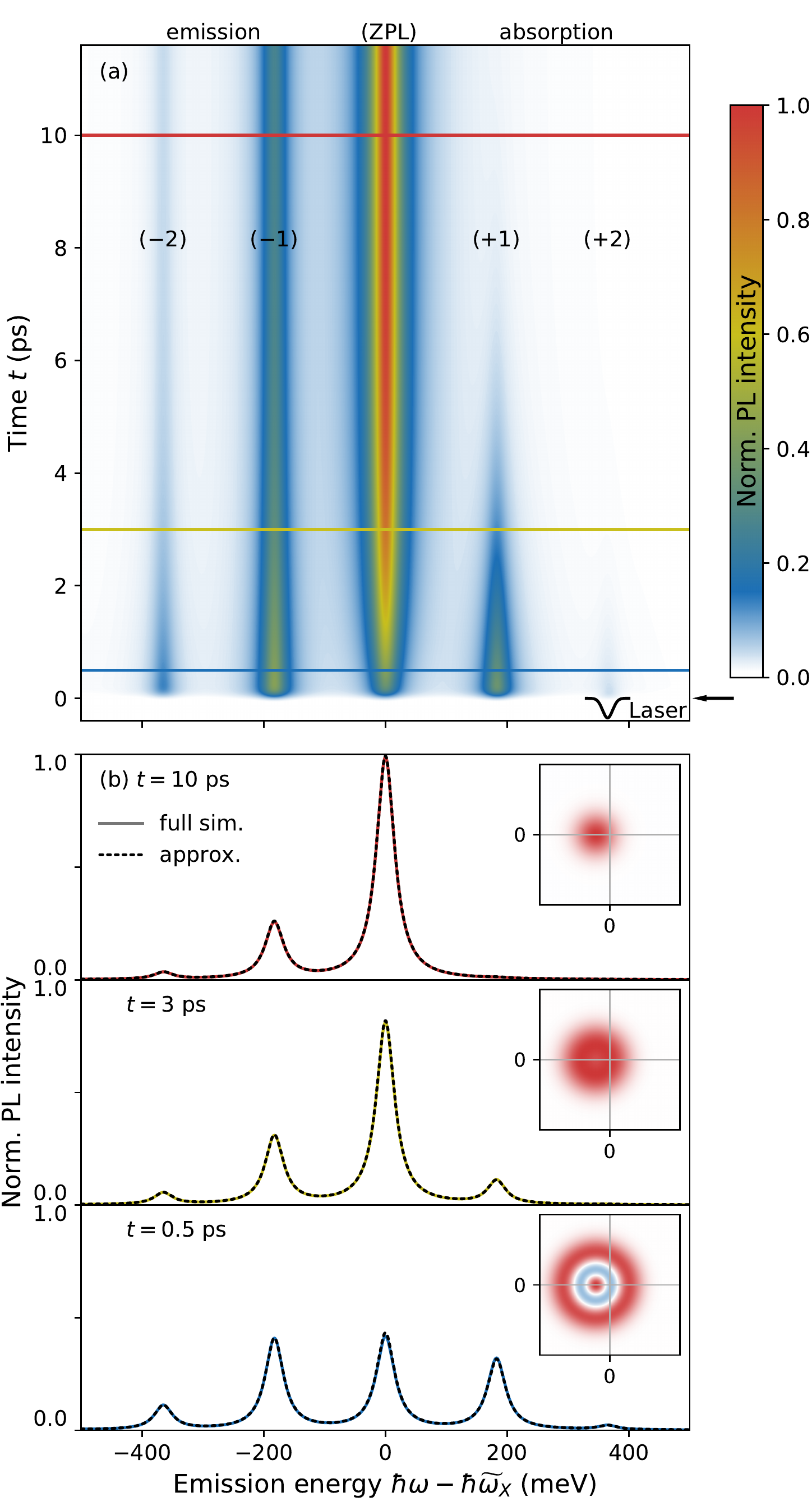}
	\caption{PL dynamics at $T=300$~K for a single LO mode after excitation on the LO-PSB(+2). (a) PL intensity color coded as a function of emission energy and time. The laser used for excitation is spectrally marked by the Gaussian - inverted for better visibility of the PSB$(+2)$ - and temporally by the arrow. (b) Cuts in (a) at different times as marked by the respectively colored horizontal line. The dashed lines show the semi-analytical calculations, employing Eq.~\eref{eq:analytical_corr}. The insets show the respective Wigner functions of the phonon state in the $\left|X\right>\left<X\right|$ subspace, blue colors show negative and red colors positive values.}
	\label{fig:1}
\end{figure}\\
For this {\it reduced model} we consider an excitation with a pulse that is detuned by $2\hbar\omega_{\rm LO}$ above the ZPL and restrict ourselves to a temperature of $T=300$~K. The simulated PL intensity is shown in Fig.~\ref{fig:1}(a) color coded as a function of photon emission energy relative to the ZPL $\hbar\omega - \hbar\tilde{\omega}_X$ and of the time $t$. The most prominent vertical line is the ZPL at energy $\hbar\omega-\hbar\tilde{\omega}_X=0$ with additional lines showing phonon emission sidebands ($-l$) at smaller and absorption sidebands ($+l$) at higher energies, where $l$ marks the number of phonons involved. The laser pulse is spectrally indicated by the black Gaussian at the bottom of the plot and temporally by the arrow on the right.\\
As explicitly shown in the SM, this detuned excitation drives the phonon subsystem associated with the excited state of the emitter into the state $B_-\left|2\right>$, where $\left|2\right>$ is the two-phonon Fock state of the single LO mode. This phonon state is reached with a high fidelity when dephasing and decay, as described by the dissipators in Eq.~\eref{eq:Lindblad}, can be neglected during the laser pulse interaction. For the TLS decay time of a few nanoseconds this is definitely fulfilled. But the faster TLS dephasing at room temperature and the phonon decay lead to a less pure phonon state and addmixtures of $B_-\left|1\right>$ and $B_-\left|0\right>$ are present after the pulse. However, for short times after the pulse in Fig.~\ref{fig:1}(a) the PSB~(+2) is slightly visible, proving that the absorption of two phonons is possible. Due to the decay of the phonons this PSB vanishes within under 2~ps. The same holds for the PSB~(+1) which survives significantly longer and is visible until almost 10~ps after the pulse. The reason for the faster decay of the PSB~(+2) is, that the transition from the state $\left|X\right>B_-\left|2\right>$ to the state $\left|X\right>B_-\left|1\right>$ via the phonon dissipator from Eq.~\eref{eq:D_LO} is double as fast as the transition from $\left|X\right>B_-\left|1\right>$ to the displaced vacuum $\left|X\right>B_-\left|0\right>$. The probability to lose a phonon is doubled, when there are double as many phonons present.\\
In contrast to the absorption side, the PSBs on the phonon emission side do not disappear for long times although they initially shrink in amplitude. The reason is that spontaneous and stimulated emission processes contribute to these signals. So as long as phonons are present in the system, stimulated emission makes these processes more likely. When the LO mode is fully decayed into the equilibrium $B_-\left|0\right>$, only spontaneous emission takes place. While all PSBs lose in amplitude, the ZPL increases for two reasons. On the one hand the decay of the excited state happens on a much longer timescale than depicted here and can therefore not be resolved. On the other hand the total PL intensity at a given time, i.e., the spectrally integrated signal of Eq.~\eref{eq:spectrum} only depends on the occupation of the TLS, integrated over the spectrometer's response time $\Gamma^{-1}$. Hence it is constant after the laser pulse on the timescale depicted in Fig.~\ref{fig:1}. This can be seen by integrating the PL spectrum in Eq.~\eref{eq:spectrum} over frequency $\omega$, forcing $\tau=0$. Thus a reduction of the PSBs has to lead to an increase of the ZPL.\\
To develop a more quantitative picture in Fig.~\ref{fig:1}(b) we show PL spectra at the specific times marked in (a) by horizontal lines with colors corresponding to the ones in (b). Additionally to each cut an inset depicts the Wigner function of the phonon state associated with the excited state of the TLS, i.e., in the $\left|X\right>\left<X\right|$ subspace. Focusing first on $t=0.5$~ps after the pulse (bottom, blue) we see that the phonon absorption sidebands have almost the same height as the emission sidebands. This shows that stimulated processes are more likely than spontaneous ones. The depicted Wigner function has two clear nodes demonstrating that a majority of the phonon occupation is indeed in the state $B_-\left|2\right>$, making the PSB~(+2) clearly visible. Moving to a larger time at $t=3$~ps (center, yellow) the PSB~(+2) has vanished and the Wigner function shows that already a significant part of the phonon occupation has reached the displaced vacuum state. For a dominating contribution of the state $B_-\left|1\right>$ the Wigner function would go to negative values in the middle of the depicted circular distribution. But we find that it only slightly drops, the reason for this being the dominating contribution from the displaced vacuum $B_-\left|0\right>$. For long times at $t=10$~ps (top, red) the phonon absorption sidebands have entirely vanished because the phonons have reached the displaced vacuum state $B_-\left|0\right>$ as confirmed by the Gaussian Wigner function. After these 10~ps a quasi-stationary limit is reached whose shape does not change anymore. Only the amplitude of the entire spectrum shrinks further for larger times due to the decay of the excited state of the TLS on a nanosecond timescale. If the signal from Eq.~\eref{eq:spectrum} is integrated over time, yielding the typical stationary PL spectrum, it will agree very well with the quasi-stationary limit, since the initial nonequilibrium dynamics of the phonons on a picosecond timescale are negligible compared to the nanosecond timescale of the emitter lifetime.\\
To understand the PSB dynamics in a slightly more quantitative manner, we take a closer look at the correlation function $\left<X^\dagger(t+\tau)X(t)\right>$ for times $t$ later than the pulse. As explained in detail in the SM, only the phonon part of the density matrix associated with the excited state, i.e., $\left<X\right|\rho(t)\left|X\right>\left|X\right>\left<X\right|$, contributes to the correlation function. We finally obtain an analytical expression that depends on the phonon configuration. For the moment we neglect TLS and LO decay between times $t$ and $t+\tau$, as they are on a nanosecond and a picosecond timescale, respectively, while the timescale for the delay $\tau$ is given by $\Gamma^{-1}$, which is below 100~fs. We will here only give the expressions for the phonon occupations, i.e., for the case that $\left<X\right|\rho(t)\left|X\right>=\sum_m f(t)c_{m,m}(t) B_-\left|m\right>\left<m\right|B_+$ with $c_{m,m}(t)$ being the occupation of the respective displaced Fock state. $f(t)=\Tr\left[X^{\dagger}X\rho(t)\right]$ is the occupation of the TLS, which is taken to be constant after the laser pulse on the picosecond timescale depicted in Fig.~\ref{fig:1}. Some comments on phonon coherences, i.e., non-diagonal elements of the phonons' density matrix will be given below. The correlation function associated with the phonon subsystem being in a displaced Fock state $B_-\left|m\right>\left<m\right|B_+$ at a time $t_0<t$ after the pulse is given by
\begin{eqnarray}\label{eq:analytical_corr}
	&&\left<X^{\dagger}(t+\tau)X(t)\right>_{B_-\left|m\right>\left<m\right|B_+}\nonumber\\
	&&=e^{-\tau/T_{\mathrm{pd}}}f(t) e^{i\tilde{\omega}_X\tau}\exp\left[-\frac{|g_{\rm LO}|^2}{\omega_{\rm LO}^2}\left(1-e^{-i\omega_{\rm LO}\tau}\right)\right]\nonumber\\
	&&\qquad \times\sum_k \frac{|g_{\rm LO}|^{2k}}{\omega_{\rm LO}^{2k}}\frac{[2\cos(\omega_{\rm LO}\tau)-2]^k}{k!}{m\choose k}\,,
\end{eqnarray}
as derived in the SM, where the LO decay is neglected between $t_0$ and $t$. We see, that the sum vanishes  for $m=0$ and we are left with the result from Ref.~\cite{wigger2019phonon}. Expanding the exponential function into a series we see that only terms $\sim e^{i(\tilde{\omega}_x-j\omega_{\rm LO})\tau}$ ($j=0,1,2,\dots$) appear. Therefore no frequencies larger than $\tilde{\omega}_X$ are possible, meaning that there is no absorption from the phononic vacuum. At $m>0$, the $\cos$-term also leads to absorption PSBs. We also see that the correlation function is independent of $t$ because the $\left|X\right>B_-\left|m\right>$ are eigenstates of $H_0$, and therefore stationary in the absence of decay and laser pulses.\\
Coherences like $B_-\left|m\right>\left<n\right|B_+$ lead to a correlation function, that oscillates in $t$ with the frequency $(m-n)\omega_{\rm LO}$, as discussed in the SM. Thus, coherences between the eigenstates of $H_0$ in the subspace where $\left|X\right>$ is occupied lead to time-dependences of the PL spectrum, in addition to the LO decay. The following two arguments explain why we do not see any such modulations in Fig.~\ref{fig:1}(a): (A) the phonons are prepared in a state, where there is no coherence, as can be seen in the Wigner functions in Fig.~\ref{fig:1}(b). Coherences would lead to distributions that depend on the angle~\cite{wigger2016quantum}. But we only find ones that are symmetric around the center at $-g_{\rm LO}/\omega_{\rm LO}$. These coherences could be created by a laser pulse that has significant spectral overlap with other PSBs or the ZPL, i.e., a laser pulse of duration shorter than $2\pi/\omega_{\rm LO}\approx 23$~fs. The creation of such phonon coherences and phonon quantum beats of optical signals has been achieved in semiconductor quantum wells, coupling to a single LO mode~\cite{banyai1995exciton, steinbach1999electron, axt1999coherent}. (B) The smallest frequency of such oscillations in $t$ would be $\omega_{\rm LO}$ with a period of around 23~fs which is significantly shorter than the integration time $\Gamma^{-1}$ in Eq.~\eref{eq:spectrum}. The oscillations would at least be smeared out significantly, reducing their visibility.\\
To compare with the analytic result from Eq.~\eref{eq:analytical_corr} for a given time $t$ we insert this result into the correlation function in Eq.~\eref{eq:spectrum}. We can do so because $\left<X^{\dagger}(t)X(t-\tau)\right> = \left<X^{\dagger}(t+\tau)X(t)\right>$ when $t-\tau$ is later than the laser pulse and as long as the LO and TLS decay can be neglegted during the delay $\tau$. At a given time $t$ the spectra are added up for each $m$ according to the phonon occupation $c_{m,m}(t)$ from the numerical simulation. The results are plotted in Fig.~\ref{fig:1}(b) as dashed black lines. We find an excellent agreement for all considered times. Thus, we conclude that for each time $t$ the spectrum can be understood as a superposition of the spectra from each phonon occupation $B_-\left|m\right>\left<m\right|B_+$, where we neglect the TLS and LO decay during the integrations over $\tau$ and $t'$ in Eq.~\eref{eq:spectrum}. The reason is that the integration time of the filter $\Gamma^{-1}$ is relatively short compared to the TLS and LO phonon lifetimes.
\subsection{Two LO modes - modelling hBN PL spectra}
\begin{figure}[h]
	\centering
	\includegraphics[width=0.5\columnwidth]{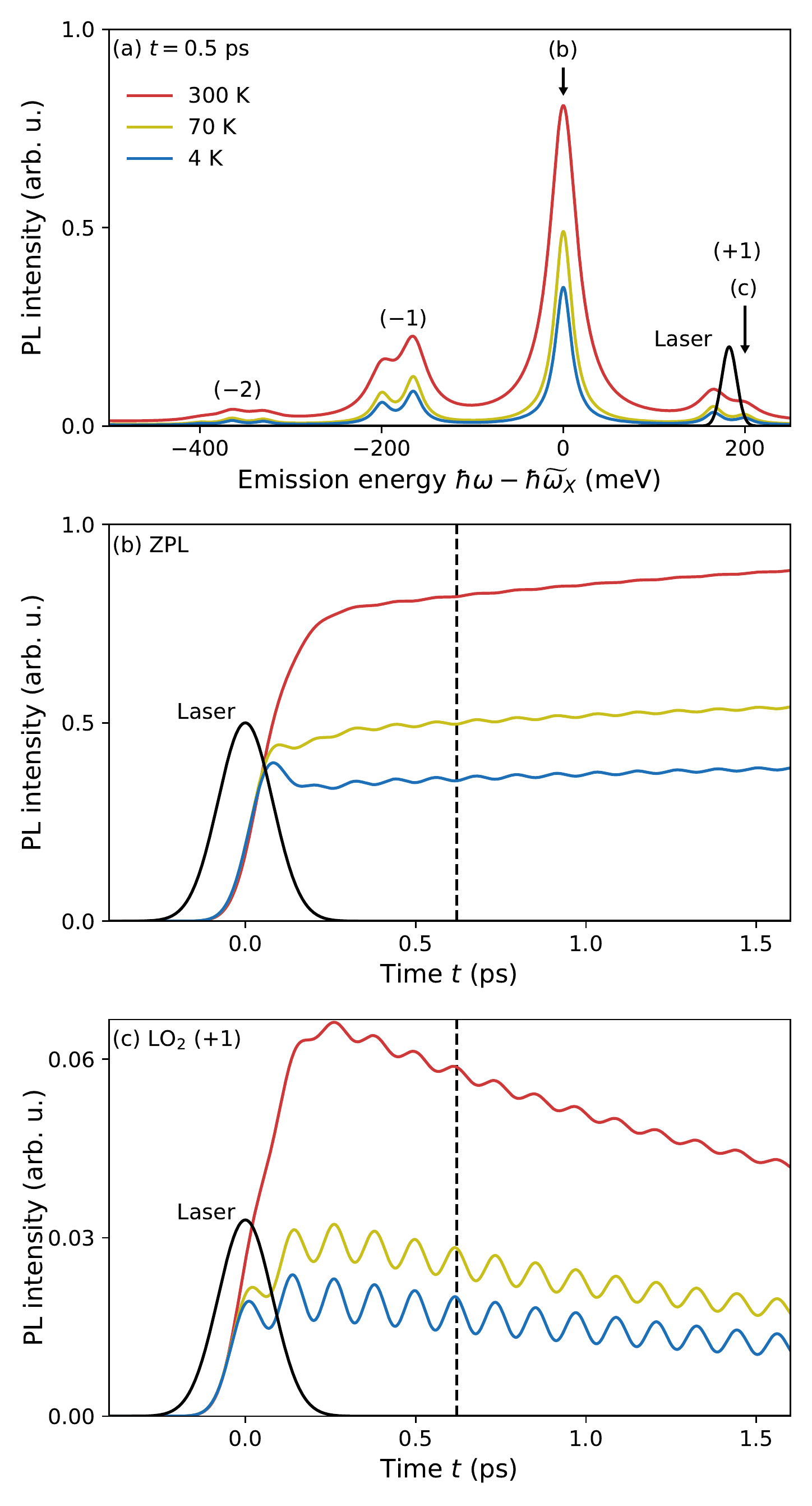}
	\caption{PL spectra for two LO modes after an excitation in the center of the PSB(+1). (a) PL spectra immediately after the laser pulse for the temperatures $T=300$ K (red), 70 K (yellow) and 4 K (blue). The spectrum of the laser pulse is depicted as a black Gaussian. (b, c) Temporal evolution of the ZPL and the LO$_2$-PSB(+1), respectively. The colors correspond to the ones in (a). The laser pulse is again the black solid line. The dashed vertical lines help to identify the anti-phased oscillation of ZPL and the PSBs.}
	\label{fig:3}
\end{figure}

Our next task is to consider a more realistic model to calculate PL spectra for the hBN emitters. We now assume two dispersionless LO modes, called LO$_1$ and LO$_2$ at energies $\hbar\omega_{{\rm LO}_1}=165$~meV and $\hbar\omega_{{\rm LO}_2}=200$~meV, respectively~\cite{wigger2019phonon}. The couplings are chosen to be $\hbar g_{{\rm LO}_1}=\hbar g_{{\rm LO}_2}=63.64$~meV, such that $\sum_{j=1,2} |g_{{\rm LO}_j}|^2/\omega_{{\rm LO}_j}^2=|g_{\rm LO}|^2/\omega_{\rm LO}^2$. By this choice the ratio between the area under the PSBs and the area under the ZPL is the same as for the single mode case discussed before. We still keep the temperature at $T=300$~K.
\begin{figure}
	\centering
	\includegraphics[width=0.5\columnwidth]{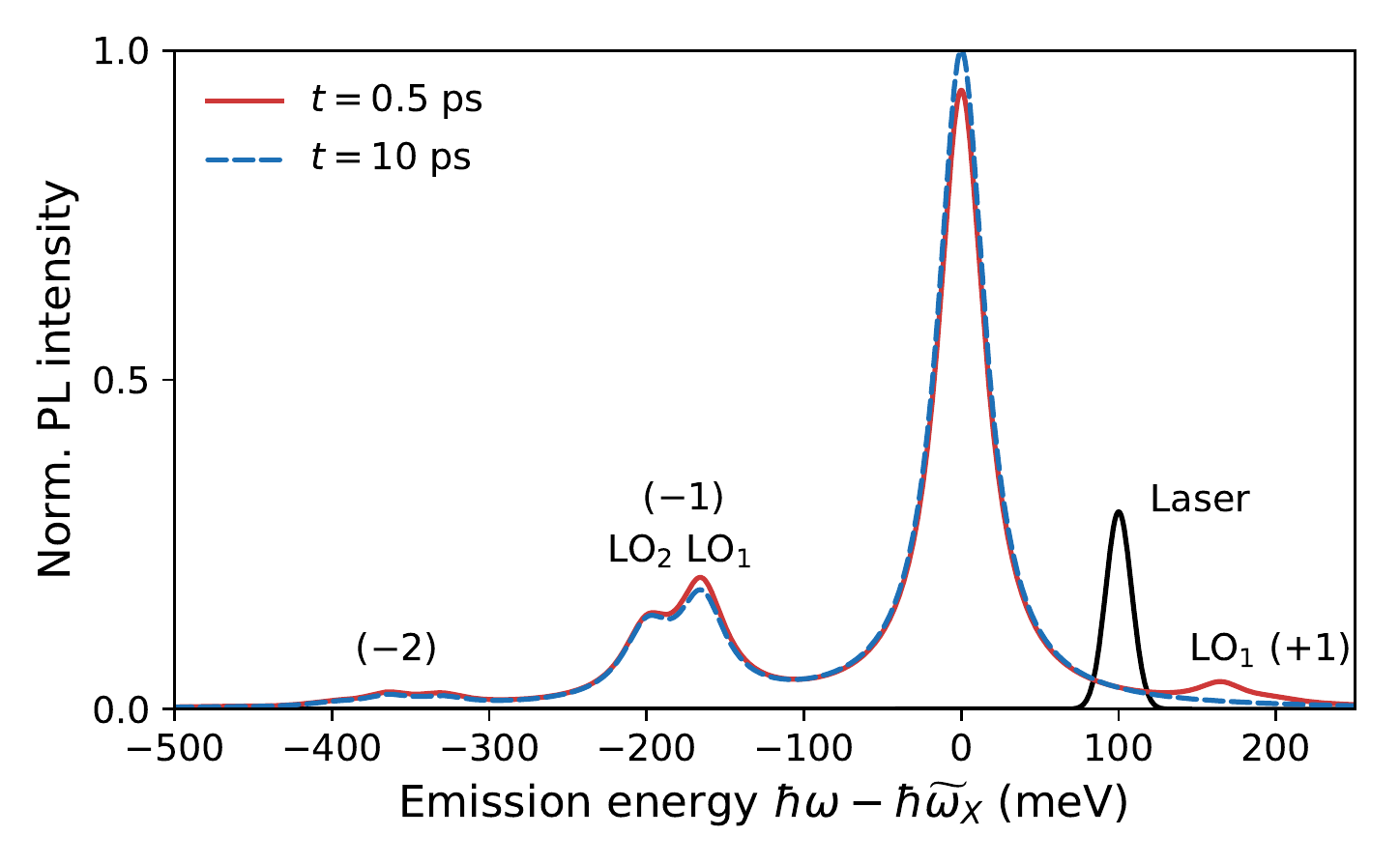}
	\caption{PL spectra for two LO modes after excitation between ZPL and first PSB at $T=300$~K. The red solid line shows the PL spectrum immediately after the pulsed excitation, the blue dashed line after the full decay of the LO phonons. The spectrum of the laser pulse is depicted as a black Gaussian.}
	\label{fig:2}
\end{figure}\\
Although the model in the previous section, where only a single LO mode was excited, was introduced as a {\it reduced model}, in the following we will demonstrate that by an appropriate optical excitation we can manage to occupy mainly one of the two modes. As depicted in Fig.~\ref{fig:2} we use a laser pulse that is detuned by 100~meV above the ZPL and therefore 65~meV below the lower LO$_1$ mode's PSB (+1). The picture shows two PL spectra, the red solid one is shortly after the laser pulse at $t=0.5$~ps and the dashed blue one at $t=10$~ps when the excited phonons are entirely decayed. We directly see that both spectra nicely resemble the double peaked PSBs ($-1$) below the ZPL well known from hBN color centers. Also the characteristic triple peak structure for the ($-2$) sidebands shows up which is also found in experiment~\cite{wigger2019phonon}. Focusing on the phonon absorption PSB~(+1) above the ZPL we only find a single peak which is the one for the LO$_1$ mode. Obviously mainly this mode gets occupied by the phonon-assisted laser pulse excitation. The LO$_2$ mode at 200~meV is not affected due to its vanishing overlap with the laser and consequently does not show up in the PSB (+1). The fact that mainly LO$_1$ is occupied shortly after the laser pulse can also be seen on the phonon emission side in the PSB ($-1$). Here we find that the red curve is slightly larger than the blue dashed one only at the LO$_1$-PSB~($-1$) energy due to the additional stimulated emission processes of this mode.\\
The naturally following question is what happens when the laser pulse is tuned exactly between the two first absorption PSBs (+1). The resulting PL spectra directly after the laser pulse at $t=0.5$~ps are plotted for three different temperatures in Fig.~\ref{fig:3}(a). The laser pulse spectrum is shown in black, the spectrum at $T=300$~K in red, at 70~K in yellow and at 4~K in blue. All curves represent typical hBN color center spectra with two additional phonon absorption peaks (+1) at 165~meV and 200~meV above the ZPL. Due to the longer dephasing times at lower temperature two features can be identified: (A) All peaks get sharper for lower temperatures and (B) the entire PL signal increases for larger temperatures. While (A) is clear, the reason for (B) lies in the different line widths. The given laser pulse spectrum has a larger overlap with the PSBs (+1) when the lines are broader. This allows for a more efficient off-resonant excitation of the TLS resulting in a stronger PL signal.
\\
While the resulting spectra are rather obvious, their dynamics provide interesting new aspects. Figures~\ref{fig:3}(b) and (c) show the temporal evolution of the ZPL and the LO$_2$-PSB~(+1), respectively, as marked in (a) by the arrows. The colors correspond to the temperatures in (a). We find that all signals oscillate in time with a period of approximately 0.1~ps. The origin of this is that in the phonon subsystem, associated with the excited state of the TLS, a superposition of the two states $B_-\left|n_{{\rm LO}_1},n_{{\rm LO}_2}\right>=B_-\left|1,0\right>$ and $B_-\left|0,1\right>$ is created. As discussed in the SM, the approximated correlation function in Eq.~\eref{eq:analytical_corr} can be extended to multiple modes. As already mentioned in the context of Eq.~\eref{eq:analytical_corr}, the coherence between the two excited phonon states $B_-\left|1,0\right> \left<0,1\right|B_+$ contributes to the PL signal. For the present case, the quantum beat oscillates with the energy difference between the two states, i.e., with $\hbar(\omega_{{\rm LO}_2}-\omega_{{\rm LO}_1})=35$~meV which corresponds to a period of 0.12~ps, long enough to be observed despite the finite time resolution of approximately $\Gamma^{-1}=66$~fs.\\
Comparing the signals for different temperatures, the visibility of the oscillations is better for lower temperatures. The reason is that the dephasing of the TLS during the interaction with the pulse results in a reduction of the purity of the phonon state, as already discussed in the context of the state preparation in the \textit{reduced model}. The same holds for the coherence resolved here, the faster dephasing at elevated temperatures results in a reduced coherence and consequently in a reduced visibility of the quantum beat. In our model we have not assumed an additional pure dephasing of the phonon modes apart from the dephasing due to their decay. Such an additional dephasing would reduce the visibility on a shorter timescale.\\
The last aspect to discuss becomes visible when comparing the dynamics of the ZPL in Fig.~\ref{fig:3}(b) with that of the PSB in (c). Note, that the intensity scale in (c) is much smaller than the one in (b). To compare the dynamics more easily, the vertical dashed lines mark the same time and appear at minima in (b) and at maxima in (c). This shows that the two signals oscillate exactly anti-phased. This effect can be understood when remembering that the full PL signal, integrated over the entire spectrum, only depends on the occupation of the excited state of the TLS, integrated over the spectrometer's temporal resolution. As the occupation of the excited state decays on a nanosecond timescale, it is virtually constant on the picosecond timescale depicted here. Therefore, when the emission in the PSBs is enhanced, the emission from the ZPL has to be reduced and vice versa. For the same reason all curves in (b) rise after the laser pulse, while the respective sideband in (c) slowly decreases. The dynamics of the remaining PSBs from Fig~\ref{fig:3}(a) is displayed in the SM for $T=4$~K, showing the in-phase oscillation of all PSBs.

\section{Conclusions}
In this work we have theoretically investigated the influence of LO phonons on time-dependent PL spectra of a hBN color center. We have first considered a \textit{reduced model} with only a single LO mode coupling to the color center. In this setting we have studied the influence of phonon decay due to anharmonicities on the PL spectra, focusing mainly on the dynamics of the PSBs. We have found excellent agreement between numerical simulations and a semi-analytical model, which allows us to gain more insight into the role of phonon occupations and phonon coherences on the dynamics of the PL spectra. In a more realistic model we have taken both LO modes with energies of 165~meV and 200~meV into account. This model leads to the characteristic two-peak PSB ($-1$) known from time-integrated PL spectra of hBN color centers~\cite{wigger2019phonon}. By using tailored optical excitation schemes we have shown that it is possible to (A) isolate one of the two modes and (B) create coherences between them. In (B) the dynamics of PL spectra exhibit quantum beats due to the 35~meV energy splitting of the two modes. A phonon quantum beat of similar frequency has been measured in semiconductor quantum wells, coupling to a single LO mode with an energy of approximately 35~meV~\cite{banyai1995exciton}. Our results demonstrate that time-dependent PL spectroscopy is a powerful tool to investigate hBN-SPEs at the same time spectrally and temporally, giving insight into the phonon dynamics of the system. 
\ack
All authors acknowledge support from the Polish National Agency for Academic Exchange (NAWA) under an APM grant. D.W. thanks NAWA for financial support within the ULAM program (No. PPN/ULM/2019/1/00064).

\section*{References}
\bibliographystyle{iopart-num}
\providecommand{\newblock}{}



\end{document}


\title[Phonon Assisted Excitation]{Controlling Photoluminescence Spectra of hBN Color Centers by Selective Phonon-Assisted Excitation: A Theoretical Proposal (Supplementary material)}

\author{Daniel~Groll$^{1}$, Thilo~Hahn$^1$, Pawe\l{}~Machnikowski$^2$, Daniel~Wigger$^{2}$ and Tilmann~Kuhn$^1$}

\address{$^1$Institute of Solid State Theory, University of M\"{u}nster, 48149 M\"{u}nster, Germany}
\address{$^2$Department of Theoretical Physics, Wroc\l{}aw University of Science and Technology, 50-370 Wroc\l{}aw, Poland }

\ead{daniel.groll@uni-muenster.de, tilmann.kuhn@uni-muenster.de}

\section{Hamiltonian and polaron transformation}
As described in the main text, we want to model the hexagonal Boron Nitride (hBN) color center as a two-level system (TLS), coupled to longitudinal optical (LO) phonons via the independent boson Hamiltonian~\cite{mahan2013many,krummheuer2002the, alicki2004pure,roszak2006path,nazir2016modelling}. The full Hamiltonian of our system, including excitation by laser pulses, is given by
\begin{eqnarray}\label{eq:H}
H&=&\underbrace{\hbar\omega_X X^{\dagger}X+\overbrace{\sum_j \hbar\omega_j^{} b_j^{\dagger}b_j^{}}^{H_{\mathrm{LO}}}+X^{\dagger}X\sum_j \left(\hbar g_j^{}b_j^{\dagger}+\hbar g_j^*b_j^{}\right)}_{H_0}+\underbrace{\frac{1}{2}\left[\mathcal{E}(t)^*X+\mathcal{E}(t)X^{\dagger}\right]}_{H_I}\,.
\end{eqnarray}
$X=\left|G\right>\left<X\right|$ is the transition operator from the excited state $\left|X\right>$ to the ground state $\left|G\right>$ with an energy splitting $\hbar\omega_X$. $b_j^{}$ is the bosonic annihilator of the $j$-th LO phonon mode with frequency $\omega_j^{}$ and coupling constant $g_j^{}$. $H_0$ is the usual independent boson model Hamiltonian, describing the coupling of the phonons to the TLS, where $H_{\mathrm{LO}}$ describes the free evolution of the LO modes. $H_I$ is the interaction of the TLS with a classical light field in the usual dipole and rotating wave approximation, modeled by the function $\mathcal{E}(t)$.\\
$H_0$ can be diagonalized, using the unitary polaron transform
\begin{equation}
\exp(S)=\exp\left[X^{\dagger}X\sum_{j}\frac{1}{\omega_j^{}}\left(g_j^{}b_j^{\dagger}-g_j^*b_j^{}\right)\right]=\left|G\right>\left<G\right|+\left|X\right>\left<X\right|B_+\,,
\end{equation}
where
\begin{equation}
B_{\pm}=\exp\left[\pm\sum_j \frac{1}{\omega_j^{}}\left(g_j^{}b_j^{\dagger}-g_j^*b_j^{}\right)\right]=\prod_j D_j\left(\pm \frac{g_j^{}}{\omega_j^{}}\right)
\end{equation}
are products of displacement operators $D_j$ for each phonon mode with displacement amplitudes $\pm g_j^{}/\omega_j^{}$~\cite{glauber1963coherent}. The polaron transformation of $H_0$ is
\begin{equation}
H_P=e^SH_0e^{-S}=\hbar\tilde{\omega}_XX^{\dagger}X+H_{\mathrm{LO}}\,,
\end{equation}
where
\begin{equation}
\tilde{\omega}_X=\omega_X-\sum_j\frac{|g_j^{}|^2}{\omega_j^{}}
\end{equation}
is the polaron shifted TLS frequency. The free time evolution operator of the independent boson model can now be easily obtained to
\begin{eqnarray}
U_0(t,t_0)&=&\exp\left[-iH_0(t-t_0)/\hbar\right]=e^{-S}\exp\left[-iH_P(t-t_0)/\hbar\right]e^S\nonumber\\
&=&e^{-S}\underbrace{\exp\left[-iH_{\mathrm{LO}}(t-t_0)/\hbar\right]}_{U_{\mathrm{LO}}(t,t_0)}\exp\left[-i\tilde{\omega}_XX^{\dagger}X(t-t_0)\right]e^S\nonumber\\
&=&\left|G\right>\left<G\right|U_{\mathrm{LO}}(t,t_0)+\left|X\right>\left<X\right|e^{-i\tilde{\omega}_X(t-t_0)}B_-U_{\mathrm{LO}}(t,t_0)B_+\,,\label{eq:U_0}
\end{eqnarray}
where $U_{\rm LO}(t,t_0)$ is the free time evolution operator of the LO modes. Decay and dephasing of the TLS are described by two phenomenological Lindblad dissipators
\begin{equation}
\mathcal{D}_{\mathrm{xd}}(\rho)=\frac{1}{T_{\mathrm{xd}}}\left(X\rho X^{\dagger}-\frac{1}{2}\lbrace X^{\dagger}X,\rho\rbrace\right)
\end{equation}
and 
\begin{equation}
\mathcal{D}_{\mathrm{pd}}(\rho)=\frac{2}{T_{\mathrm{pd}}}\left(X^{\dagger}X\rho X^{\dagger}X-\frac{1}{2}\lbrace X^{\dagger}X,\rho\rbrace\right)\,,
\end{equation}
respectively, where $T_{\mathrm{xd}}$ is the lifetime of the TLS and $T_{\mathrm{pd}}$ its pure dephasing time. Up until now, leaving out the phonon decay which is derived in the next section, the Lindblad equation describing the time evolution of the density matrix $\rho$ of the system reads
\begin{equation}
\frac{\mathrm{d}}{\mathrm{d}t}\rho(t)=-\frac{i}{\hbar}\left[H_0+H_I(t),\rho(t)\right]+\mathcal{D}_{\mathrm{xd}}[\rho(t)]+\mathcal{D}_{\mathrm{pd}}[\rho(t)]\,.
\end{equation}
\section{Derivation of the phonon dissipator from phonon anharmonicity}
In the following we derive the dissipator for the LO phonons, describing their decay. To this aim, we consider the anharmonic coupling of the LO phonons to low energy modes like acoustic phonons. From the third order anharmonic coupling we include only the terms that describe creation or annihilation of LO phonons $\sim b_j^{\dagger}, b_j^{}$. In absence of the laser pulse, the full Hamiltonian including the anharmonic coupling is given by
\begin{equation}
\tilde{H}=H_0+\sum_n\hbar\Omega_na_n^{\dagger}a_n+\sum_{j,m,n}\left(\hbar V_{jmn}b^{}_ja_m^{\dagger}a_n^{\dagger}+h.c.\right)\,,
\end{equation}
where $a_m^{\dagger}$ is the creation operator for low energy phonons of mode $m$. Due to the anharmonic coupling, we now include processes, where a LO phonon is annihilated and at the same time two low energy phonons are created, as described by the interaction term $b_j^{}a_m^{\dagger}a_n^{\dagger}$. The reverse processes are of course also included. The matrix elements $V_{jmn}$ are the coupling constants for these processes and can be chosen symmetric between $m$ and $n$. To stay consistent with the simulations presented in the main text, we neglect coupling of the low energy modes with frequencies $\Omega_n$ and associated bosonic operators $a_n^{(\dagger)}$ to the TLS. To derive the dissipator, we go to the polaron frame for the LO modes, such that
\begin{eqnarray}
	\tilde{H}^P&=&e^S\tilde{H}e^{-S}=\tilde{H}_S+\tilde{H}_B+\tilde{H}_{SB}\,,\\
	\tilde{H}_S&=&e^SH_0e^{-S}=\hbar\tilde{\omega}_XX^{\dagger}X+H_{\mathrm{LO}}\,,\\
	\tilde{H}_B&=&\sum_n\hbar\Omega_n a_n^{\dagger}a_n^{}\,,\\
	\tilde{H}_{SB}&=&\sum_{j,m,n}\left[\hbar V_{jmn}\left(b_j^{}-\frac{g_j^{}}{\omega_j^{}}X^{\dagger}X\right)a_m^{\dagger}a_n^{\dagger}+h.c.\right]\,.
\end{eqnarray}
Here, we introduced the separation of the full Hamiltonian into system, bath and system-bath interaction part, called $\tilde{H}_S$, $\tilde{H}_B$ and $\tilde{H}_{SB}$, respectively. Note that the average of $\tilde{H}_{SB}$ vanishes, if we assume a thermal state for the low energy phonons. In the derivation of the Lindblad equation, we follow Chapter 3 of Ref.~\cite{breuer2002theory}. First, we need to write the interaction Hamiltonian in the interaction picture with respect to $\tilde{H}_S$ and $\tilde{H}_B$ and bring it into the following form
\begin{equation}\label{eq:H_I_I}
\tilde{H}_{SB}^I(t)=\sum_{\alpha,\omega}=\exp(-i\omega t)A_{\alpha}(\omega)B_{\alpha}^I(t)\,,
\end{equation}
where the $B$ operators act only on the bath, i.e., the low energy phonons. The $A$ operators act only on the system, i.e., the TLS and the LO phonons. Direct calculation gives
\begin{equation}
\tilde{H}_{SB}^I(t)=\sum_{j,m,n}\left[\hbar V_{jmn}\left(b_j^{}e^{-i\omega_j t}-\frac{g_j^{}}{\omega_j^{}}X^{\dagger}X\right)a_m^{\dagger}e^{i\Omega_mt}a_n^{\dagger}e^{i\Omega_n t}+h.c.\right]\,.
\end{equation}
From this, we can identify
\begin{eqnarray}
	A_{1j}(\omega)&=&b_j\delta_{\omega,\omega_j}-\frac{g_j^{}}{\omega_j^{}}X^{\dagger}X\delta_{\omega,0}\,,\\
	A_{2j}(\omega)&=&b_j^{\dagger}\delta_{\omega,-\omega_j}-\frac{g_j^*}{\omega_j}X^{\dagger}X\delta_{\omega,0}\,,\\
	B_{1j}^I(t)&=&\sum_{m,n}\hbar V_{jmn}a_m^{\dagger}e^{i\Omega_mt}a_n^{\dagger}e^{i\Omega_n t}\,,\\
	B_{2j}^I(t)&=&\sum_{m,n}\hbar V_{jmn}^*a_me^{-i\Omega_mt}a_ne^{-i\Omega_n t}\,,
\end{eqnarray}
where we wrote the index $\alpha$ from Eq.~\eref{eq:H_I_I} as a double index $kj$ with $k=1,2$. In the next step, we calculate the bath correlation functions
\begin{equation}
G_{\alpha\beta}(\tau)=\frac{1}{\hbar^2}\left<\left[B_{\alpha}^I(\tau)\right]^{\dagger}B_{\beta}^I(0)\right>\,.
\end{equation}
Assuming a thermal state for the low energy phonons, we see that $G_{1i,2j}=G_{2i,1j}=0$. We further get
\begin{eqnarray}
	G_{2i,2j}(\tau)&=&\sum_{m,n,m',n'}V_{imn}V_{jm'n'}^*e^{i(\Omega_m+\Omega_n)\tau}\left<a_m^{\dagger}a_n^{\dagger}a_{m'}a_{n'}\right>\nonumber\\
	&=&2\sum_{m,n}V_{imn}V_{jmn}^*e^{i(\Omega_m+\Omega_n)\tau}n(\Omega_m)n(\Omega_n)
\end{eqnarray}
with $n(\Omega)$ being the thermal occupation of phonons of frequency $\Omega$. In the derivation we used the symmetry of the coupling matrix $V_{jmn}$ with respect to the indices $m$ and $n$. In the same fashion we arrive at
\begin{equation}
G_{1i,1j}(\tau)=2\sum_{m,n}V_{imn}V_{jmn}^*e^{-i(\Omega_m+\Omega_n)\tau}\left[n(\Omega_m)+1\right]\left[n(\Omega_n)+1\right]\,.
\end{equation}
From the correlation functions we can derive the decay rates and renormalization parameters by performing first the one-sided Fourier transform, to obtain
\begin{eqnarray}
	\Gamma_{1i,1j}(\omega)&=&\int\limits_0^{\infty}\mathrm{d}\tau e^{i\omega\tau}G_{1i,1j}(\tau)\\
	&=&2\sum_{m,n}V_{imn}V_{jmn}^*\left[n(\Omega_m)+1\right]\left[n(\Omega_n)+1\right]\nonumber\\
	&&\left[\pi\delta(\omega-\Omega_m-\Omega_n)+iP\frac{1}{\omega-\Omega_m-\Omega_n}\right]\,,\nonumber
\end{eqnarray}
\begin{eqnarray}
	\Gamma_{2i,2j}(\omega)&=&\int\limits_0^{\infty}\mathrm{d}\tau e^{i\omega\tau}G_{2i,2j}(\tau)\\
	&=&2\sum_{m,n}V_{imn}V_{jmn}^*n(\Omega_m)n(\Omega_n)\nonumber\\
	&&\left[\pi\delta(\omega+\Omega_m+\Omega_n)+iP\frac{1}{\omega+\Omega_m+\Omega_n}\right]\nonumber\,.
\end{eqnarray}
Assuming that the conditions for the Born-Markov approximations hold (fast decay of the bath correlation functions and weak coupling between bath and system), we need to verify that the secular approximation holds, to arrive at a true Lindblad equation. The timescale, which is induced by the interaction with the bath, i.e., the decay timescale is on the order of a few picoseconds~\cite{cusco2018isotopic}. The intrinsic timescale of the system is here $\max_{\omega\neq\omega'}2\pi|\omega-\omega'|^{-1}$ with $\omega,\,\,\omega'=0,\omega_1,\omega_2,..\,$. In the case of the \textit{reduced model} with a single LO mode from the main text, this timescale is $2\pi\hbar (182.5\mathrm{ meV})^{-1}\approx 20$~fs. In the case of the more realistic model with two LO modes the intrinsic timescale is $2\pi\hbar (200\mathrm{ meV}-165\mathrm{ meV})^{-1}\approx 100$~fs. Both of these values are well below the picosecond-timescale of the decay. Thus we can apply the secular approximation, which is necessary to arrive at the Lindblad form. For a continuum of modes, like in the case of acoustic phonons, the dissipator takes a more complicated form than the one derived here, since the secular approximation does not hold in that case.\\
The Lamb shift Hamiltonian, which describes renormalization of the system energies, is here of the form
\begin{eqnarray}
H_{LS}&=&\sum_{\omega,\alpha,\beta}\frac{\hbar}{2i}[\Gamma_{\alpha\beta}(\omega)-\Gamma_{\beta\alpha}^*(\omega)]A_{\alpha}^{\dagger}(\omega)A_{\beta}(\omega)\nonumber\\
&=&\Delta E_X X^{\dagger}X+\sum_j\hbar\Delta\omega_j b_j^{\dagger}b_j+\mathrm{const.}
\end{eqnarray}
with $\Delta E_X$ being the energy shift of the TLS and $\Delta\omega_j$ being the frequency shift of the LO phonons. Due to the structure of the $A_{\alpha}$'s, this Hamiltonian simply shifts the energies of the polaron and the optical phonons and we can simply absorb these shifts into the phonon frequencies and the TLS energy later.\\
The dissipator in the polaron frame, which describes decay processes induced by the bath, is given by
\begin{eqnarray}
	\mathcal{D}_P(\rho^P)&=&\sum_{\omega,\alpha,\beta}\left[\Gamma_{\alpha\beta}(\omega)+\Gamma_{\beta\alpha}^*(\omega)\right]\left[A_{\beta}(\omega)\rho^P A_{\alpha}^{\dagger}(\omega)-\frac{1}{2}\lbrace A_{\alpha}^{\dagger}(\omega)A_{\beta}(\omega),\rho^P\rbrace\right]\nonumber\\
	&=&\tilde{\gamma}_{\mathrm{pd}}\left[X^{\dagger}X\rho^P X^{\dagger}X-\frac{1}{2}\lbrace X^{\dagger}X,\rho^P\rbrace\right]\nonumber\\
	&+&2\sum_j\mathrm{Re}\left[\Gamma_{1j,1j}(\omega_j)\right]\left[b_j\rho^P b_j^{\dagger}-\frac{1}{2}\lbrace b_j^{\dagger}b_j,\rho^P\rbrace\right]\nonumber\\
	&+&2\sum_j\mathrm{Re}\left[\Gamma_{2j,2j}(-\omega_j)\right]\left[b_j^{\dagger}\rho^P b_j-\frac{1}{2}\lbrace b_jb_j^{\dagger},\rho^P\rbrace\right]\,,
\end{eqnarray}
where $\rho^P=\exp(S)\rho\exp(-S)$ is the density matrix of the LO+TLS system in the polaron frame. The first term describes dephasing of the polaron with the rate $\tilde{\gamma}_{\mathrm{pd}}$ and stems from the zero frequency components of the $A_{\alpha}(\omega)$, the second one describes decay of the LO phonons by spontaneous and induced emission of low energy phonons and the last term describes creation of LO phonons by absorption of low energy phonons. In the derivation we assumed that all LO modes have different frequencies. By looking at the respective rates for phonon decay and creation, i.e., $\mathrm{Re}\left[\Gamma_{1j,1j}(\omega_j)\right]$ and $\mathrm{Re}\left[\Gamma_{2j,2j}(-\omega_j)\right]$, we see that absorption and emission processes have to fulfill the energy conservation condition $\omega_j=\Omega_m+\Omega_n$. As an example decay process we consider one, where the two low energy phonons, that are emitted, have the same energy. At room temperature we have $k_BT\approx25$\,meV and thus for $\hbar\omega_j>160$meV we have $n(\omega_j/2)\ll 1$. Thus we can safely ignore this kind of absorption and induced emission of phonons. Assuming that the energy of the LO phonon is distributed almost equally among the two low energy phonons for all decay processes, the dissipator can be written as
\begin{equation}
\mathcal{D}_P(\rho^P)=\tilde{\gamma}_{\mathrm{pd}}\left[X^{\dagger}X\rho^P X^{\dagger}X-\frac{1}{2}\lbrace X^{\dagger}X,\rho^P\rbrace\right]+\sum_j\gamma_j\left[b_j\rho^P b_j^{\dagger}-\frac{1}{2}\lbrace b_j^{\dagger}b_j,\rho^P\rbrace\right]\,.
\end{equation}
$\tilde{\gamma}_{\mathrm{pd}}$ is a correction to the phenomenological pure dephasing and will later be absorbed into the dephasing dissipator $\mathcal{D}_{\mathrm{pd}}$, $\gamma_j$ is the decay rate of the $j$-th LO mode. The full time evolution for the density matrix in the polaron frame $\rho^P$ is given by
\begin{equation}
\frac{\mathrm{d}}{\mathrm{d}t}\rho^P(t)=-\frac{i}{\hbar}\left[(\hbar\tilde{\omega}_X+\Delta E_X)X^{\dagger}X+\sum_j\hbar(\omega_j+\Delta\omega_j)b_j^{\dagger}b_j,\rho^P(t)\right]+\mathcal{D}_P\left[\rho^P(t)\right]\,.
\end{equation}
In the next step we go back to the original frame with the unitary operator
\begin{equation}
e^S=\exp\left[X^{\dagger}X\sum_j\frac{1}{\omega_j}\left(g_jb_j^{\dagger}-g_j^*b_j\right)\right]=\exp\left[X^{\dagger}X\sum_j\frac{1}{\omega_j+\Delta\omega_j}\left(\tilde{g}_jb_j^{\dagger}-\tilde{g}_j^*b_j\right)\right]\,
\end{equation}
with
\begin{equation}
\tilde{g_j}=g_j\frac{\omega_j+\Delta\omega_j}{\omega_j}\,.
\end{equation}
The time evolution in this frame is given by
\begin{eqnarray}
\frac{\mathrm{d}}{\mathrm{d}t}\rho&=&\frac{\mathrm{d}}{\mathrm{d}t}e^{-S}\rho^Pe^S=-\frac{i}{\hbar}\left[\left(\hbar\tilde{\omega}_X+\Delta E_X+\sum_j\hbar\frac{|\tilde{g}_j|^2}{\omega_j+\Delta\omega_j}\right)X^{\dagger}X\right.\nonumber\\
&&\left.+\sum_j\hbar(\omega_j+\Delta\omega_j)b_j^{\dagger}b_j+X^{\dagger}X\sum_j\hbar\left(\tilde{g}_jb^{\dagger}+\tilde{g}_j^*b_j\right),\rho\right]+\mathcal{D}(\rho)
\end{eqnarray}
with 
\begin{eqnarray}
	\mathcal{D}(\rho)&=&\tilde{\gamma}_{\mathrm{pd}}\left[X^{\dagger}X\rho X^{\dagger}X-\frac{1}{2}\lbrace X^{\dagger}X,\rho\rbrace\right]\nonumber\\
	&+&\sum_j\underbrace{\gamma_j\left[\left|G\right>\left<G\right|b_j\rho b_j^{\dagger}\left|G\right>\left<G\right|-\frac{1}{2}\lbrace\left|G\right>\left<G\right|b_j^{\dagger}b_j,\rho\rbrace\right]}_{\mathcal{D}_{{\rm LO}_j}^{G}(\rho)}\nonumber\\
	&+&\sum_j\underbrace{\gamma_j\left[\left|X\right>\left<X\right|B_-b_jB_+\rho B_-b_j^{\dagger}B_+\left|X\right>\left<X\right|-\frac{1}{2}\lbrace\left|X\right>\left<X\right|B_-b_j^{\dagger}b_jB_+,\rho\rbrace\right]}_{\mathcal{D}_{{\rm LO}_j}^{X}(\rho)}\nonumber\\
	&+&\sum_j\underbrace{\gamma_j \left[\left|G\right>\left<G\right|b_j\rho B_-b_j^{\dagger}B_+\left|X\right>\left<X\right|+\left|X\right>\left<X\right|B_-b_jB_+\rho b_j^{\dagger}\left|G\right>\left<G\right|\right]}_{\mathcal{D}_{{\rm LO}_j}^{XG}(\rho)}\,.
\end{eqnarray}
Thus, we arrive at the form of the dissipator for the phonons, as presented in the main text in Eq.~(5). Absorbing all renormalizations due to the interaction with the low energy phonons into the original parameters and adding the phenomenological dissipators and the laser driving, the full time evolution of the system is given by (see also Eq.~(6) of the main text)
\begin{equation}\label{eq:Lindblad}
\frac{\rm d}{{\rm d}t}\rho(t)=-\frac{i}{\hbar}\left[ H_0+H_I(t),\rho(t)\right]+\mathcal{D}_{\mathrm{xd}}[\rho(t)]+\mathcal{D}_{\mathrm{pd}}[\rho(t)]+\sum_j \mathcal{D}_{\mathrm{LO}_j}[\rho(t)]\,,
\end{equation}
where the LO phonon dissipator was written as
\begin{eqnarray}
\mathcal{D}_{{\rm LO}_j}(\rho)&=& \mathcal{D}_{{\rm LO}_j}^{G}(\rho)+\mathcal{D}_{{\rm LO}_j}^{X}(\rho)+\mathcal{D}_{{\rm LO}_j}^{XG}(\rho) \label{eq:D_LO}\,.
\end{eqnarray}
Note, that in the derivation of the phonon dissipator we neglected the laser pulse. Since the decay is on a timescale of a few picoseconds and the laser pulse duration is approximately $100$~fs, we assume that the decay is largely unaffected by the laser pulse interaction.
\section{Derivation of the formula for the time-dependent spectrum}
In Ref.~\cite{eberly1977time} the concept of time-dependent spectroscopy is theoretically described. There, the time-dependent spectrum is given in terms of the first order correlation functions of the quantized light field at the position of a Fabry-Perot interferometer, filtering the light prior to detection. The correlation functions of the light field can be expressed in terms of operators of the TLS~\cite{mollow1969power}, such that the full form of the time-dependent spectrum for our model is
\begin{equation}
S(t,\omega;\Gamma)=\Gamma^2\int\limits_{-\infty}^t\mathrm{d} t_1\,\int\limits_{-\infty}^t\mathrm{d} t_2\,\underbrace{e^{-(\Gamma-i\omega)(t-t_1)}e^{-(\Gamma+i\omega)(t-t_2)}\left<X^{\dagger}(t_1)X(t_2)\right>}_{s(t_1,t_2,\omega;\Gamma)}\,,
\end{equation}
where $\omega$ is the frequency of the detected light, $t$ the time and $\Gamma$ the spectral width of the interferometer. We will now time-order the integration
\begin{eqnarray}
S(t,\omega;\Gamma)&=&\Gamma^2\left[\int\limits_{-\infty}^t\mathrm{d} t_1\,\int\limits_{-\infty}^{t_1}\mathrm{d} t_2\,+\int\limits_{-\infty}^t\mathrm{d} t_1\,\int\limits_{t_1}^t\mathrm{d} t_2\,\right]s(t_1,t_2,\omega;\Gamma)\nonumber\\
&=&\Gamma^2\left[\int\limits_{-\infty}^t\mathrm{d} t_1\,\int\limits_{-\infty}^{t_1}\mathrm{d} t_2\,+\int\limits_{-\infty}^t\mathrm{d} t_2\,\int\limits_{-\infty}^{t_2}\mathrm{d} t_1\,\right]s(t_1,t_2,\omega;\Gamma)\nonumber\\
&=&2\Gamma^2\mathrm{Re}\int\limits_{-\infty}^t\mathrm{d} t_1\,\int\limits_{-\infty}^{t_1}\mathrm{d} t_2\,e^{-(\Gamma-i\omega)(t-t_1)}e^{-(\Gamma+i\omega)(t-t_2)}\left<X^{\dagger}(t_1)X(t_2)\right>\,.
\end{eqnarray}
When we divide the last expression for $S(t,\omega;\Gamma)$ by $e^{-2\Gamma t}$, we see, that the integrand does not depend on $t$ anymore. Calculating the derivative of $S(t,\omega;\Gamma) e^{2\Gamma t}$ with respect to $t$ amounts to replacing all $t_1$ appearing in the integrand by $t$ and removing the $t_1$-integration. With these steps we find (with the substitution $\tau=t-t_2$) that
\begin{eqnarray}
\frac{\mathrm{d}}{\mathrm{d}t}\frac{S(t,\omega;\Gamma)}{e^{-2\Gamma t}}&=&2\Gamma^2\mathrm{Re} \int\limits_0^{\infty}\mathrm{d}\tau\,e^{-\Gamma\tau}e^{2\Gamma t}e^{-i\omega\tau}\left<X^{\dagger}(t)X(t-\tau)\right>\,.
\end{eqnarray}
Integrating over $t$, we find for the time-dependent spectrum
\begin{equation}
S(t,\omega;\Gamma)=2\Gamma^2\mathrm{Re}\int\limits_{0}^{\infty}\mathrm{d}\tau\,\int\limits_{-\infty}^t\mathrm{d} t'e^{-\Gamma\tau}e^{-2\Gamma (t-t')}e^{-i\omega\tau}\left<X^{\dagger}(t')X(t'-\tau)\right>\, ,
\end{equation}
where we assumed that the correlation function vanishes in the far past, i.e., $S(-\infty,\omega;\Gamma)=0$. This is the form of the time-dependent spectrum, as used in the main text in Eq.~(7).
\section{State preparation}
In this section we discuss the state preparation of the phonon subsystem after laser excitation. Since the lifetime $T_{\mathrm{xd}}$ of the emitter is on a nanosecond-timescale but we are interested in sub-picosecond dynamics, we can safely neglect the dissipator $\mathcal{D}_{\mathrm{xd}}$, that describes TLS decay, when we are interested in the phonon state shortly after the laser pulse.\\
To keep the discussion as simple as possible, we will here also neglect the phonon decay, as described by the dissipators $\mathcal{D}_{{\rm LO}_j}$ from Eq.~\eref{eq:D_LO}. Their influence on the state preparation is discussed in a qualitative manner at the end of this section.\\
We begin our discussion with the equation of motion
\begin{equation}\label{eq:state_prep_unitary}
\frac{\rm d}{{\rm d}t}\rho_{\rm tot}(t)=-\frac{i}{\hbar}\left[H_0+H_{\mathrm{pd}}+H_I(t),\rho_{\rm tot}(t)\right]\,,
\end{equation}
where $H_{\mathrm{pd}}$ is the Hamiltonian that describes the coupling of the system to environmental degrees of freedom that induce dephasing. The coupling shall be such, that the Born-Markov approximations hold, i.e., the total density matrix $\rho_{\rm tot}(t)\approx\rho(t)\otimes\rho_{\mathrm{pd}}$ factorizes into $\rho(t)$, which describes our system, and $\rho_{\mathrm{pd}}$, which describes the environment. This coupling shall lead to the TLS dephasing dissipator $\mathcal{D}_{\mathrm{pd}}$, such that the above equation of motion is approximately equivalent to the Lindblad equation~\cite{breuer2002theory}
\begin{equation}\label{eq:state_prep_lindblad}
\frac{\rm d}{{\rm d}t}\rho(t)=-\frac{i}{\hbar}\left[H_0+H_I(t),\rho(t)\right]+\mathcal{D}_{\mathrm{pd}}[\rho(t)]\,.
\end{equation}
Starting in the distant past with the state $\rho(-\infty)=\left|G\right>\left|\lbrace 0\rbrace\right>\left<\lbrace 0\rbrace\right|\left<G\right|$, where $\left|\lbrace 0\rbrace \right>$ is the multi-mode vacuum of the LO modes, our goal is to derive the phonon state associated with the excited state of the TLS, i.e., $\left<X\right|\rho(t)\left|X\right>$ at a time $t$ after the laser pulse.\\
Assuming a weak laser pulse, such that we can perform the calculations in the lowest order of perturbation theory, we can write
\begin{equation}
\left<X\right|\rho(t)\left|X\right>=\Tr_{\mathrm{pd}}\left[ \left<X\right|U(t,-\infty)\rho(-\infty)\rho_{\mathrm{pd}} U^{\dagger}(t,-\infty)\left|X\right>\right]\,,
\end{equation}
where $\Tr_{\mathrm{pd}}$ denotes the trace over the environmental degrees of freedom. Now we insert a perturbative expansion in terms of the interaction $H_I$ of the full time evolution operator, i.e.,
\begin{eqnarray}
U(t,t_0)&=&\hat{T}\exp\bigg\lbrace-\frac{i}{\hbar}\int\limits_{t_0}^t\mathrm{d}t'\left[H_0+H_{\mathrm{pd}}+H_I(t')\right]\bigg\rbrace\nonumber\\
&\approx&U_{0,\mathrm{pd}}(t,t_0) -\frac{i}{\hbar}\int\limits_{t_0}^t \mathrm{d}t'\,U_{0,\mathrm{pd}}(t,t')H_I(t')U_{0,\mathrm{pd}}(t',t_0)\,.
\end{eqnarray}
Here, $\hat{T}$ is the time-ordering operator and 
\begin{equation}
U_{0,\mathrm{pd}}(t,t_0)=\exp\left[-i(H_0+H_{\mathrm{pd}})(t-t_0)/\hbar\right]\,
\end{equation}
is the time evolution operator of the system and environment in the absence of the laser pulse. Now we can write the phonon state as
\begin{eqnarray}\label{eq:state_1}
\left<X\right|\rho(t)\left|X\right>=\Tr_{\mathrm{pd}}\Bigg[\int\limits_{-\infty}^t\mathrm{d}t_1\int\limits_{-\infty}^t\mathrm{d}t_2\,&&\left<X\right|U_{0,\mathrm{pd}}(t,t_1)H_I(t_1)U_{0,\mathrm{pd}}(t_1,-\infty)\left|G\right>\left|\lbrace 0\rbrace\right>\rho_{\mathrm{pd}}\nonumber\\
&&\left<\lbrace 0\rbrace\right|\left<G\right|U_{0,\mathrm{pd}}^{\dagger}(t_2,-\infty)H_I(t_2)U_{0,\mathrm{pd}}^{\dagger}(t,t_2)\left|X\right>\Bigg]\,.
\end{eqnarray}
To evaluate this expression, we need some properties of $U_{0,\mathrm{pd}}(t,t_0)$. First of all, it can be used to find the solutions of Eq.~\eref{eq:state_prep_unitary} or, equivalently, Eq.~\eref{eq:state_prep_lindblad} in absence of the laser pulse. Since $H_0$ and the dephasing do not change the occupation of the TLS, we find, given an arbitary phonon state $\rho_B(t_0)$
\begin{eqnarray}
U_{0,\mathrm{pd}}(t,t_0)\left|G\right>\left<G\right| \rho_B(t_0)\rho_{\mathrm{pd}}U_{0,\mathrm{pd}}^{\dagger}(t,t_0)&=&U_0(t,t_0)\left|G\right>\left<G\right| \rho_B(t_0)U_0^{\dagger}(t,t_0)\rho_{\mathrm{pd}}\,,\nonumber\\
U_{0,\mathrm{pd}}(t,t_0)\left|X\right>\left<X\right| \rho_B(t_0)\rho_{\mathrm{pd}}U_{0,\mathrm{pd}}^{\dagger}(t,t_0)&=&U_0(t,t_0)\left|X\right>\left<X\right| \rho_B(t_0)U_0^{\dagger}(t,t_0)\rho_{\mathrm{pd}}\,,\label{eq:U_02}\\
U_{0,\mathrm{pd}}(t,t_0)\left|X\right>\left<G\right| \rho_B(t_0)\rho_{\mathrm{pd}}U_{0,\mathrm{pd}}^{\dagger}(t,t_0)&=&U_0(t,t_0)\left|X\right>\left<G\right| \rho_B(t_0)U_0^{\dagger}(t,t_0)e^{-\frac{t-t_0}{T_{\mathrm{pd}}}}\rho_{\mathrm{pd}}\,,\nonumber
\end{eqnarray}
where $U_0$ is the free time evolution operator associated with $H_0$, see Eq.~\eref{eq:U_0}. These relations can be derived by solving Eq.~\eref{eq:state_prep_lindblad} in the absence of laser pulses for the initial condition $\rho(t_0)=\rho_{\rm TLS}\rho_B(t_0)$ by observing $\mathcal{D}_{\mathrm{pd}}(\left|X\right>\left<X\right|)=\mathcal{D}_{\mathrm{pd}}(\left|G\right>\left<G\right|)=0$ and $\mathcal{D}_{\mathrm{pd}}(\left|X\right>\left<G\right|)=-\frac{1}{T_{\mathrm{pd}}}\left|X\right>\left<G\right|$. Here $\rho_{\rm TLS}$ stands for one of the three TLS density matrix components from Eqs.~\eref{eq:U_02}.\\
Now we can procede solving Eq.~\eref{eq:state_1} by observing that up until time $\min(t_1,t_2)$, the density matrix component that is evolved in time is $\left|G\right>\left<G\right|$. From time $\min(t_1,t_2)$ to $\max(t_1,t_2)$ it is either $\left|G\right>\left<X\right|$ or $\left|X\right>\left<G\right|$ and from that time on it is $\left|X\right>\left<X\right|$. Thus we can write
\begin{eqnarray}
\left<X\right|\rho(t)\left|X\right>=\int\limits_{-\infty}^t\mathrm{d}t_1\int\limits_{-\infty}^t\mathrm{d}t_2\,e^{-\frac{|t_2-t_1|}{T_{\mathrm{pd}}}}&&\left<X\right|U_{0}(t,t_1)H_I(t_1)U_{0}(t_1,-\infty)\left|G\right>\left|\lbrace 0\rbrace\right>\nonumber\\
&&\left<\lbrace 0\rbrace\right|\left<G\right|U_{0}^{\dagger}(t_2,-\infty)H_I(t_2)U_{0}^{\dagger}(t,t_2)\left|X\right>\,.
\end{eqnarray}
Using our formula for the time evolution operator $U_0$ from Eq.~\eref{eq:U_0} and the full expression for the interaction Hamiltonian $H_I$ from Eq.~\eref{eq:H}, we find
\begin{eqnarray}\label{eq:state_prep_general}
\left<X\right|\rho(t)\left|X\right>&=&\frac{1}{4}\int\limits_{-\infty}^t\mathrm{d}t_1\int\limits_{-\infty}^t\mathrm{d}t_2\,\mathcal{E}(t_1)\mathcal{E}^*(t_2)e^{-\frac{|t_2-t_1|}{T_{\mathrm{pd}}}}e^{-i\tilde{\omega}_X(t_2-t_1)}\nonumber\\
&&\times B_-U_{\rm LO}(t,t_1)B_+\left|\lbrace 0\rbrace\right>\left<\lbrace 0\rbrace\right|B_-U_{\rm LO}^{\dagger}(t,t_2)B_+\nonumber\\
&=&\frac{1}{4}\int\limits_{t_0}^t\mathrm{d}t_1\int\limits_{t_0}^t\mathrm{d}t_2\,\mathcal{E}(t_1)\mathcal{E}^*(t_2)e^{-\frac{|t_2-t_1|}{T_{\mathrm{pd}}}}e^{-i\tilde{\omega}_X(t_2-t_1)}\nonumber \\
&&\times B_-U_{\rm LO}(t,t_0)B_+(t_1)\left|\lbrace 0\rbrace\right>\left<\lbrace 0\rbrace\right|B_-(t_2)U_{\rm LO}^{\dagger}(t,t_0)B_+\,.
\end{eqnarray}
Here, $B_{\pm}(t)=U_{\rm LO}(t,t_0)^{\dagger}B_{\pm}U_{\rm LO}(t,t_0)$ is the multi-mode displacement operator in the interaction picture with respect to the free LO phonon Hamiltonian and $t_0$ is some arbitrary initial time before the laser pulse interaction.
\subsection{Single phonon mode}
In the case of the \textit{reduced model}, where we only consider a single phonon mode with frequency $\omega_{\rm LO}$ and coupling $g_{\rm LO}$, the multi-mode displacement operators $B_{\pm}$ reduce to single-mode displacement operators. Specifically, in the interaction picture they are given by
\begin{equation}
B_{+}(t)=D\left(+\frac{g_{\rm LO}}{\omega_{\rm LO}}e^{i\omega_{\rm LO}(t-t_0)}\right)=\sum_m\hat{O}_m^{+}e^{im\omega_{\rm LO}(t-t_0)}\,,\quad B_-(t)=B_+(t)^{\dagger}\,,
\end{equation}
where $D$ is the displacement operator of the LO mode and we used a Fourier series representation. The operator $\hat{O}^{+}_m$ creates $m$ phonons, for $m$ being positive and annihilates $m$ phonons for $m$ being negative. This can be seen by using the reciprocal Fourier representation
\begin{equation}
\hat{O}^+_m=\frac{\omega_{\rm LO}}{2\pi}\int\limits_0^{\frac{2\pi}{\omega_{\rm LO}}}\mathrm{d} t\,B_+(t)e^{-im\omega_{\rm LO} (t-t_0)}\,.
\end{equation}
From this, we can explicitly show its role as creator of $m$ phonons
\begin{eqnarray}
\left<k\right|\hat{O}^+_m\left|l\right>&=&\frac{\omega_{\rm LO}}{2\pi}\int\limits_0^{\frac{2\pi}{\omega_{\rm LO}}}\mathrm{d} t\,\left<k\right|U_{\rm LO}^{\dagger}(t,t_0)B_+U_{\rm LO}(t,t_0)\left|l\right>e^{-im\omega_{\rm LO} (t-t_0)}\nonumber\\
&=&\frac{\omega_{\rm LO}}{2\pi}\int\limits_0^{\frac{2\pi}{\omega_{\rm LO}}}\mathrm{d} t\,\left<k\right|B_+\left|l\right>e^{-i(m+l-k)\omega_{\rm LO} (t-t_0)}=\left<k\right|B_+\left|l\right>\delta_{m,k-l}\,.
\end{eqnarray}
Using the Fourier representation, we can write Eq.~\eref{eq:state_prep_general} as
\begin{eqnarray}
\left<X\right|\rho(t)\left|X\right>&=&\sum_{m,n}\frac{1}{4}\int\limits_{t_0}^t\mathrm{d}t_1\int\limits_{t_0}^t\mathrm{d}t_2\,\mathcal{E}(t_1)\mathcal{E}^*(t_2)e^{-\frac{|t_2-t_1|}{T_{\mathrm{pd}}}}e^{im\omega_{\rm LO}(t_1-t_0)}e^{-in\omega_{\rm LO}(t_2-t_0)}\nonumber\\
&&\times e^{-i\tilde{\omega}_X(t_2-t_1)}B_-U_{\rm LO}(t,t_0)\hat{O}_m^+\left| 0\right>\left< 0\right|\hat{O}_n^{+,\dagger}U_{\rm LO}^{\dagger}(t,t_0)B_+\,.
\end{eqnarray}
Here, $\left|0\right>$ is the vacuum of the single LO mode. In the main text, in the chapter on the \textit{reduced model}, the laser is tuned to the phonon sideband (PSB)~$(+2)$ and the evolution of the phonon state is observed by time-dependent spectroscopy. To discuss this state preparation in detail, let us separate the carrier frequency $\omega_l$ from the laser field, i.e., $\mathcal{E}(t)=\tilde{\mathcal{E}}(t)e^{-i\omega_l (t-t_0)}$, where $\tilde{\mathcal{E}}$ is the envelope of the laser field, which has a constant complex phase. Tuning the laser to the PSB~$(+2)$ means, that we set $\omega_l=\tilde{\omega}_X+2\omega_{\rm LO}$. With this, we obtain
\begin{eqnarray}
\left<X\right|\rho(t)\left|X\right>&=&\sum_{m,n}\frac{1}{4}\int\limits_{t_0}^t\mathrm{d}t_1\int\limits_{t_0}^t\mathrm{d}t_2\,\tilde{\mathcal{E}}(t_1)\tilde{\mathcal{E}}^*(t_2)e^{-\frac{|t_2-t_1|}{T_{\mathrm{pd}}}}e^{i(m-2)\omega_{\rm LO}(t_1-t_0)}e^{-i(n-2)\omega_{\rm LO}(t_2-t_0)}\nonumber\\
&&\times B_-U_{\rm LO}(t,t_0)\hat{O}_m^+\left| 0\right>\left< 0\right|\hat{O}_n^{+,\dagger}U_{\rm LO}^{\dagger}(t,t_0)B_+\,.
\end{eqnarray}
For simplicity, we will consider a rectangular shaped pulse of length $\Delta t$, centered around time $0$, such that the phonon state after the pulse at $t>\Delta t/2$ becomes
\begin{eqnarray}\label{eq:state_prep_final}
\left<X\right|\rho(t)\left|X\right>&=&\sum_{m,n}\frac{1}{4\Delta t^2}\int\limits_{-\Delta t/2}^{\Delta t/2}\mathrm{d}t_1\int\limits_{-\Delta t/2}^{\Delta t/2}\mathrm{d}t_2\,e^{-\frac{|t_2-t_1|}{T_{\mathrm{pd}}}}e^{i(m-2)\omega_{\rm LO}(t_1-t_0)}e^{-i(n-2)\omega_{\rm LO}(t_2-t_0)}\nonumber\\
&&\times B_-U_{\rm LO}(t,t_0)\hat{O}_m^+\left| 0\right>\left< 0\right|\hat{O}_n^{+,\dagger}U_{\rm LO}^{\dagger}(t,t_0)B_+\,.
\end{eqnarray}
The double integral can be evaluated to yield
\begin{eqnarray}\label{eq:double_int}
\int\limits_{-\Delta t/2}^{\Delta t/2}\mathrm{d}t_1&&\int\limits_{-\Delta t/2}^{\Delta t/2}\mathrm{d}t_2\,e^{-\frac{|t_2-t_1|}{T_{\mathrm{pd}}}}e^{i(m-2)\omega_{\rm LO}t_1}e^{-i(n-2)\omega_{\rm LO}t_2}\nonumber\\
&&=\frac{2\sin\left[(m-n)\omega_{\rm LO}\Delta t/2\right]}{(m-n)\omega_{\rm LO}}\frac{\frac{2}{T_{\mathrm{pd}}}}{(n-2)^2\omega_{\rm LO}^2+\frac{1}{T_{\mathrm{pd}}^2}}\nonumber\\
&&+2\mathrm{Re}\left\lbrace\frac{e^{i(m-n)\omega_{\rm LO}\Delta t/2}-e^{-\Delta t/T_{\mathrm{pd}}}e^{-i(m+n-4)\omega_{\rm LO}\Delta t/2}}{\left[-i(n-2)\omega_{\rm LO}+\frac{1}{T_{\mathrm{pd}}}\right]\left[i(m-2)\omega_{\rm LO}-\frac{1}{T_{\mathrm{pd}}}\right]}\right\rbrace\,.
\end{eqnarray}
We see, that the first term grows with $\Delta t$, whereas the second term is bounded for all $\Delta t$, since
\begin{eqnarray}\label{eq:upper_bound}
\mathrm{Re}&&\left\lbrace\frac{e^{i(m-n)\omega_{\rm LO}\Delta t/2}-e^{-\Delta t/T_{\mathrm{pd}}}e^{-i(m+n-4)\omega_{\rm LO}\Delta t/2}}{\left[-i(n-2)\omega_{\rm LO}+\frac{1}{T_{\mathrm{pd}}}\right]\left[i(m-2)\omega_{\rm LO}-\frac{1}{T_{\mathrm{pd}}}\right]}\right\rbrace\nonumber\\
&&\leq \left|\frac{e^{i(m-n)\omega_{\rm LO}\Delta t/2}-e^{-\Delta t/T_{\mathrm{pd}}}e^{-i(m+n-4)\omega_{\rm LO}\Delta t/2}}{\left[-i(n-2)\omega_{\rm LO}+\frac{1}{T_{\mathrm{pd}}}\right]\left[i(m-2)\omega_{\rm LO}-\frac{1}{T_{\mathrm{pd}}}\right]}\right|\nonumber\\
&&\leq 2\left|\left[-i(n-2)\omega_{\rm LO}+\frac{1}{T_{\mathrm{pd}}}\right]\right|^{-1}\left|\left[i(m-2)\omega_{\rm LO}-\frac{1}{T_{\mathrm{pd}}}\right]\right|^{-1}\nonumber\\
&&=2\sqrt{\left[(n-2)^2\omega_{\rm LO}^2+T_{\mathrm{pd}}^{-2}\right]\left[(m-2)^2\omega_{\rm LO}^2+T_{\mathrm{pd}}^{-2}\right]}^{-1}\nonumber\\
&&\approx 2\sqrt{\left[(n-2)^2\omega_{\rm LO}^2\right]\left[(m-2)^2\omega_{\rm LO}^2\right]}^{-1}\sim\omega_{\rm LO}^{-2}\,.
\end{eqnarray}
In the last line we approximated the upper bound by using the fact, that for the parameters relevant for this paper we have $T_{\mathrm{pd}}\omega_{\rm LO}\gg 1$. In the limit of $\Delta t\rightarrow \infty$ and $T_{\mathrm{pd}}\rightarrow \infty$, i.e., cw-laser excitation and vanishing dephasing, the Lorentzians and the sinc in Eqs.~\eref{eq:double_int} and \eref{eq:upper_bound} force the phonons to be prepared in a displaced two-phonon state, i.e., $m=n=2$
\begin{eqnarray}
\left<X\right|\rho(\infty)\left|X\right>&\sim& B_-U_{\rm LO}(\infty,-\infty)\hat{O}_2^+\left|0\right>\left<0\right|\hat{O}_2^{+,\dagger}U_{\rm LO}^{\dagger}(\infty,-\infty)B_+\nonumber\\
&\sim& B_-\left|2\right>\left<2\right|B_+\,.
\end{eqnarray}
A finite dephasing time $T_{\mathrm{pd}}$ leads to a contribution from displaced Fock states with a Fock state occupation different from $2$. If, however, $\Delta t$ is long enough, such that $\omega_{\rm LO}\Delta t\gg 1$ at the same time, the sinc in the first term of Eq.~\eref{eq:double_int} assures $m=n$. All contributions from the second term  in Eq.~\eref{eq:double_int} with $m\neq 2$ or $n\neq 2$ are suppressed in Eq.~\eref{eq:state_prep_final} in this limit, such that the phonon state in Eq.~\eref{eq:state_prep_final} contains no coherences like $B_-\left|n\right>\left<m\right|B_+$, but only occupations like $B_-\left|n\right>\left<n\right|B_+$. The condition $\omega_{\rm LO}\Delta t\gg 1$ is easily satisfied for a $\Delta t=100$~fs pulse duration and a frequency of $\hbar\omega_{\rm LO}=182.5$~meV, yielding $\omega_{\rm LO}\Delta t\approx 27.7\gg 1$. In the main text, a gaussian pulse of 80~fs standard deviation was used, which has a temporal width close to the rectangle pulse, used here. Thus, for the parameters used in the paper, we expect a phonon state, that is a mixture of occupations of the form $B_-\left|n\right>\left<n\right|B_+$, and that the most weight is on the $n=2$ state. We see, that larger dephasing leads to a less pure phonon state, as the contributions to the density matrix go from $n=2$ towards a broader distribution, the larger $T_{\mathrm{pd}}$ gets. Phonon decay during the pulse will shift the weight of the distribution of states towards smaller $n$, as eventually the ground state $B_-\left|0\right>$ is reached. 
\section{Correlation functions for semi-analytical spectra}
In this section we derive the analytic expressions for the correlation function $\left<X^{\dagger}(t+\tau)X(t)\right>$ for the case, that $\tau>0$ and $t$ is a time later than the laser pulse interaction. $\tau$ shall be small enough, such that we can neglect phonon decay and TLS decay. We include dephasing, such that the equations of motion, describing our system are given by Eqs.~\eref{eq:state_prep_unitary} or, equivalently, \eref{eq:state_prep_lindblad}.\\
At time $t$, the state of the system shall be prepared in $\rho(t)=f(t)\left|X\right>\left<X\right|\otimes \rho_B(t)$, where $\rho_B(t)$ is an arbitary phonon state and $f(t)\leq 1$ is the occupation of the TLS. Since the independent boson model Hamiltonian, as well as the dephasing, do not change the occupation of the TLS, the only contribution to the correlation function is from the subspace of states, where the TLS is in the excited state $\left|X\right>$. That is why we restrict ourselves to the form of $\rho(t)$ as given above. All other contributions from TLS configurations different to $\left|X\right>\left<X\right|$ would vanish anyway. This is true even when we include TLS and phonon decay, since both do not create contributions to the $\left|X\right>\left<X\right|$ subspace over time. The TLS decay just reduces the occupation exponentially. Thus, the subspace of states, where the TLS is in the excited state, takes the prominent role in the calculation of the correlation functions for times later than the laser pulse excitation. This is mentioned in the main text in the context of Eq.~(7).\\
Keeping in mind the discussion in the context of Eqs.~\eref{eq:U_02} and the time evolution operator $U_0$ from Eq.~\eref{eq:U_0}, we can give the following expression for the correlation function
\begin{eqnarray}\label{eq:corr_func}
\left<X^{\dagger}(t+\tau)X(t)\right>&=&f(t)e^{-\frac{\tau}{T_{\mathrm{pd}}}}\Tr\left[U_0^{\dagger}(t+\tau,t)X^{\dagger}U_0(t+\tau,t)X\rho_B(t)\right]\nonumber\\
&=&f(t)e^{-\frac{\tau}{T_{\mathrm{pd}}}}e^{i\tilde{\omega}_X\tau}\Tr\left[B_-U_{\rm LO}^{\dagger}(t+\tau,t)B_+X^{\dagger}U_{\rm LO}(t+\tau,t)X\rho_B(t)\right]\nonumber\\
&=&f(t)e^{-\frac{\tau}{T_{\mathrm{pd}}}}e^{i\tilde{\omega}_X\tau}\Tr\left[B_-B_+(\tau)\rho_B(t)\right]\,.
\end{eqnarray}
In the last step we used, that the multi-mode displacement operator in the interaction picture with respect to the free LO Hamiltonian depends only on the time difference, i.e.,
\begin{equation}
U_{\rm LO}^{\dagger}(t+\tau,t)B_+U_{\rm LO}(t+\tau,t)\equiv B_+(\tau)
\end{equation}
does not depend on $t$. The only $t$-dependence enters through a potential time dependence of the phonon state $\rho_B$ associated with the subspace, in which the TLS is in the excited state. The state at a time $t_0$ prior to $t$, but later than the laser, is given by $\rho(t_0)=f(t)\left|X\right>\left<X\right|\otimes \rho_B(t_0)$, where we again made use of the fact, that we neglect the TLS decay and that TLS dephasing does not change the TLS occupation $f$. $\rho_B(t)$ and $\rho_B(t_0)$ are related through $U_0(t,t_0)$, such that
\begin{equation}\label{eq:phonon_t_dependence}
\rho_B(t)=B_-U_{\rm LO}(t,t_0)B_+\rho_B(t_0) B_- U_{\rm LO}^{\dagger}(t,t_0)B_+\,.
\end{equation}
If the phonons are prepared in a mixture of displaced Fock states for each mode, i.e., a mixture of states of the form $B_-\left|n\right>\left<n\right|B_+$ in the case of a single mode, then $\rho_B$ does not depend on time. Coherences between displaced Fock states would lead to a time-dependence.
\subsection{Single phonon mode}
We consider now the case of a single phonon mode, as presented in the main text as \textit{reduced model}. The phonons are prepared in a state $\rho_B(t_0)=\sum_{m,n}c_{m,n}B_-\left|m\right>\left<n\right|B_+$, leading to a phonon state at time $t$ of $\rho_B(t)=\sum_{m,n}c_{m,n}e^{-i\omega_{\rm LO}(m-n)(t-t_0)}B_-\left|m\right>\left<n\right|B_+$. Thus, the correlation function is given by
\begin{eqnarray}\label{eq:corr_func_analytic}
\left<X^{\dagger}(t+\tau)X(t)\right>&=&\sum_{m,n}c_{m,n}e^{-i\omega_{\rm LO}(m-n)(t-t_0)} f(t)e^{-\frac{\tau}{T_{\mathrm{pd}}}}e^{i\tilde{\omega}_X\tau}\Tr\left[B_-B_+(\tau)B_-\left|m\right>\left<n\right|B_+\right]\nonumber\\
&\equiv&\sum_{m,n}c_{m,n}e^{-i\omega_{\rm LO}(m-n)(t-t_0)} f(t)e^{-\frac{\tau}{T_{\mathrm{pd}}}}e^{i\tilde{\omega}_X\tau}G_{mn}(\tau)\nonumber\\
&\equiv&\sum_{m,n} c_{m,n}\left<X^{\dagger}(t+\tau)X(t)\right>_{B_-\left|m\right>\left<n\right|B_+}\,.
\end{eqnarray}
All we need to calculate are matrix elements of displacement operators
\begin{eqnarray}\label{eq:G_mn_tau}
G_{mn}(\tau)&=&\Tr\left[B_-B_+(\tau)B_-\left|m\right>\left<n\right|B_+\right]=\left<n\right|D\left(\frac{g_{\rm LO}}{\omega_{\rm LO}}e^{i\omega_{\rm LO}\tau}\right)D\left(-\frac{g_{\rm LO}}{\omega_{\rm LO}}\right)\left|m\right>\nonumber\\
&=&\left<n\right|D\left[\frac{g_{\rm LO}}{\omega_{\rm LO}}\left(e^{i\omega_{\rm LO}\tau}-1\right)\right]\left|m\right>\exp\left[-i\frac{|g_{\rm LO}|^2}{\omega_{\rm LO}^2}\sin(\omega_{\rm LO}\tau)\right]\,.
\end{eqnarray}
Here, $D$ denotes the single mode displacement operator. Now we make use of
\begin{equation}
\left<m\right|D(\alpha)\left|m\right>=e^{-\frac{1}{2}|\alpha|^2}\sum_k \frac{(-|\alpha|^2)^k}{k!}{m\choose k}\,,
\end{equation}
which can be shown by using the normal-ordered form of the displacement operator. We insert this formula together with Eq.~\eref{eq:G_mn_tau} into Eq.~\eref{eq:corr_func_analytic}and thus arrive at
\begin{eqnarray}\label{eq:analytical_corr}
\left<X^{\dagger}(t+\tau)X(t)\right>_{B_-\left|m\right>\left<m\right|B_+}&&=e^{-\tau/T_{\mathrm{pd}}}f(t) e^{i\tilde{\omega}_X\tau}\exp\left[-\frac{|g_{\rm LO}|^2}{\omega_{\rm LO}^2}\left(1-e^{-i\omega_{\rm LO}\tau}\right)\right]\nonumber\\
&&\qquad \times\sum_k \frac{|g_{\rm LO}|^{2k}}{\omega_{\rm LO}^{2k}}\frac{[2\cos(\omega_{\rm LO}\tau)-2]^k}{k!}{m\choose k}\,,
\end{eqnarray}
as presented in the main text in Eq.~(8). We do not present the formula for the coherences explicitly, as the interesting feature, which is the $t$-dependence, is already obvious in Eq.~\eref{eq:corr_func_analytic}. We see, that coherences generate a $t$-dependence, which oscillates with at least the frequency $\omega_{\rm LO}$ in the single mode case.
\subsection{Two phonon modes}
The generalization to multiple phonon modes is straightforward. For simplicity we will consider here only the case of the coherence between the modes LO$_1$ and LO$_2$ from the main text, which leads to a quantum beat. This can be seen by considering the case of $\rho_B(t_0)=B_-\left|0,1\right>\left<1,0\right|B_+$, where the first entry of the ket and bra is for mode LO$_1$ and the second entry for LO$_2$. Such a coherence occurs in a superposition of the states $B_-\left|0,1\right>$ and $B_-\left|1,0\right>$, which can be prepared by setting the laser frequency between the two PSB~$(+1)$, as discussed in the main text.\\
From Eq.~\eref{eq:phonon_t_dependence} we obtain the time dependence of the coherence to be $\rho_B(t)=e^{-i(\omega_{{\rm LO}_2}-\omega_{{\rm LO}_1})(t-t_0)}B_-\left|0,1\right>\left<1,0\right|B_+$. This can be inserted into Eq.~\eref{eq:corr_func} to yield the part of the correlation function that originates from this coherence. The calculations are essentially the same as in the single mode case, as the multi-mode displacement operators are just products of single-mode displacement operators for each mode and thus the trace in Eq.~\eref{eq:corr_func} factorizes. We do not present the result here, as the essential feature for the paper is the $t$-dependence of the correlation function and thus of the time-dependent spectrum, which is given by the frequency difference $\omega_{{\rm LO}_2}-\omega_{{\rm LO}_1}$ between the modes.
\newpage
\section{Quantum beat: sideband dynamics}
\begin{figure}[h]
	\centering
	\includegraphics[width=0.7\linewidth]{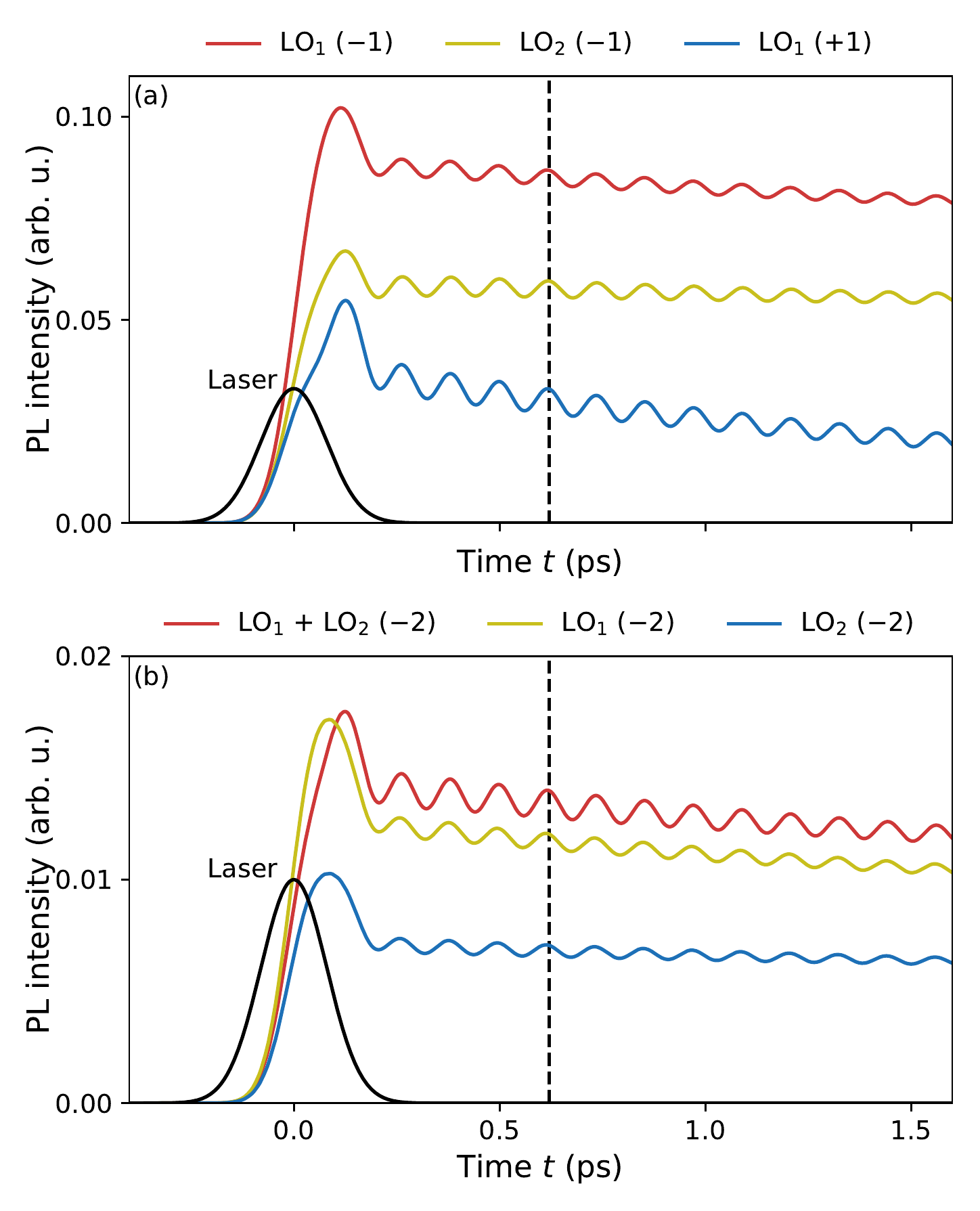}
	\caption{Temporal evolution of the PSBs after a laser excitation spectrally between the LO$_1$-PSB$(+1)$ and the LO$_2$-PSB$(+1)$. The calculations are performed at $T=4$~K and correspond to the ones from Fig.~(4) in the main text. In the main text the dynamics of the ZPL and the LO$_{2}$-PSB$(+1)$ are shown. (a) shows the dynamics of the remaining peak from the PSB$(+1)$ and of the two peaks from the PSB$(-1)$. (b) shows the dynamics of all three peaks from the PSB$(-2)$. The laser pulse is denoted by the black solid line. The dashed vertical lines help to identify the in-phase oscillation of the PSBs.}
	\label{fig:SM_1}
\end{figure}

In Fig.~(4) of the main text, phonon quantum beats of the zero phonon line (ZPL) and the LO$_2$-PSB$(+1)$ are shown after a laser pulse excitation spectrally between the LO$_1$-PSB$(+1)$ and the LO$_2$-PSB$(+1)$. In Fig.~\ref{fig:SM_1}, we show the dynamics of the remaining PSBs at $T=4$~K. Fig.~\ref{fig:SM_1}(a) shows the temporal evolution of the remaining PSB$(+1)$-peak, i.e., the LO$_1$-PSB$(+1)$, and of the two peaks from the PSB$(-1)$. Fig.~\ref{fig:SM_1}(b) shows the temporal evolution of all three peaks from the PSB$(-2)$. Both figures contain a black curve, denoting the temporal form of the laser pulse, and a dashed vertical line, which helps to identify the in-phase oscillation of all PSBs. In fact, the dashed vertical lines are at the same position as the ones from Fig.~(4) in the main text, such that we see, that all PSBs oscillate in phase, whereas the ZPL oscillates out of phase with respect to them.
\newpage
\section*{References}
\bibliographystyle{iopart-num}
\providecommand{\newblock}{}